\documentstyle[aps,prb,floats,epsf]{revtex}
\begin{document}
\advance\textheight by 0.2in

% [28] submitted to Phys. Rev. B (1998)

\draft
\twocolumn[\hsize\textwidth\columnwidth\hsize\csname@twocolumnfalse%
\endcsname
\title{Properties of the Bose glass phase in irradiated superconductors \\ 
  near the matching field}

\author{Carsten Wengel$^{1}$ and Uwe Claus T\"auber$^{2}$}

\address{$^{1}$ Institut f\"ur Theoretische Physik, Universit\"at G\"ottingen, 
  Bunsenstr. 9, D--37073 G\"ottingen, Germany \\ 
  $^{2}$ Institut f\"ur Theoretische Physik T34, Physik-Department der
  Technischen Universit\"at M\"unchen, \\
  James-Franck-Stra\ss e, D--85747 Garching, Germany} 

\date{\today}

\maketitle

\begin{abstract}
Structural and transport properties of interacting localized flux lines in the
Bose glass phase of irradiated superconductors are studied by means of Monte
Carlo simulations near the matching field $B_\Phi$, where the densities of
vortices and columnar defects are equal. 
For a completely random columnar pin distribution in the $xy$-plane transverse
to the magnetic field, our results show that the repulsive vortex interactions
destroy the Mott insulator phase which was predicted to occur at $B = B_\Phi$.
On the other hand, for ratios of the penetration depth to average defect
distance $\lambda / d \lesssim 1$, characteristic remnants of the Mott
insulator singularities remain visible in experimentally accessible quantities
as the magnetization, the bulk modulus, and the magnetization relaxation, when
$B$ is varied near $B_\Phi$. 
For spatially more regular disorder, e.g., a nearly triangular defect
distribution, we find that the Mott insulator phase can survive up to
considerably large interaction range $\lambda / d$, and may thus be observable
in experiments.
\end{abstract}

]

\section{Introduction} \label{intro_sec}

The remarkably rich phase diagram of magnetic flux lines in high-$T_c$
superconductors, especially when subject to point and/or extended disorder, has
attracted considerable experimental and theoretical interest. \cite{blatter94}
For the purpose of applying type-II superconductors in external magnetic
fields, an effective vortex pinning mechanism is essential, in order to
minimize dissipative losses caused by the Lorentz-force induced movement of
flux lines across the sample.
This issue becomes even more important for the high-$T_c$ superconductors,
because the strongly enhanced thermal fluctuations in these materials render
the Abrikosov flux lattice unstable.
Thus, well below the upper critical field $H_{\rm c_2}(T)$, there occurs a
first-order melting transition leading to a normal-conducting flux liquid
phase. \cite{nelson88}
It turns out to be rather difficult to pin such a vortex liquid consisting of
strongly fluctuating flexible lines by just intrinsic point disorder (e.g.,
oxygen vacancies), although asymptotically a truly superconducting
low-temperature vortex glass phase was predicted. \cite{fisher91}

\begin{figure}[t]
\centerline{\epsfysize 6cm \epsfbox{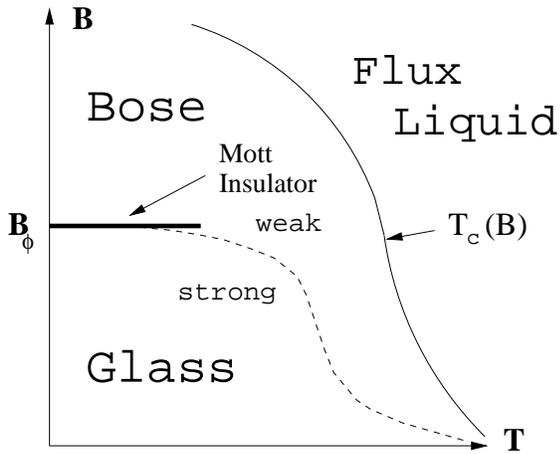}}
\caption{Phenomenological phase diagram of a type-II superconductor with strong
  correlated disorder. A sharp phase transition line $T_c(B)$ separates the
  flux liquid from the Bose glass phase. Within this phase, one may identify a
  crossover line discriminating between the weakly and strongly pinned Bose
  glass. The Mott insulator is located deep within the Bose glass at low
  temperatures, appearing as a line phase {\em precisely} at the matching field
  $B = B_\Phi$.}
\label{bg-phase_fig}
\end{figure}

Therefore, in addition to point defects, the influence of extended or
correlated disorder, promising stronger pinning effects, on vortex transport
properties has been considered.
Experimentally, linear damage tracks have been produced in oxide
superconductors by irradiation with high-energy heavy ions.   
These columnar defects serve as strong pinning centers for the flux lines, by
significantly increasing the critical current and shifting the irreversibility
line upwards (see Ref.~[\onlinecite{blatter94}], Sec.~IX~B for references).
Theoretically, the statistical mechanics problem of directed lines interacting
with columnar pins can be mapped onto the quantum mechanics of two-dimensional
bosons subject to static point disorder via the identification of the vortex
trajectories ${\bf r}_i(z)$ with particle world lines in imaginary time
\cite{nelson88,nelson89,lyuksyutov92,nelson92}.
In this mathematical analogy, thermal fluctuations $\sim k_{\rm B}T$ map onto
quantum fluctuations $\sim \hbar$, the vortex line tension ${\tilde\epsilon}_1$
represents the boson mass, and the sample thickness $L$ along the $z$ direction
parallel to the magnetic field ${\bf B}$ corresponds to the inverse temperature
$\beta \hbar$ of the quantum system (such that the thermodynamic limit 
$L \to \infty$ implies zero temperature for the fictitious bosons).
The boson picture can then be employed to immediately construct the expected
phase diagram for flux lines interacting with columnar defects, as depicted in
Fig.~\ref{bg-phase_fig}, \cite{lyuksyutov92,nelson92} and its properties can be
analyzed by utilizing the (scaling) theory of bosons subject to static point
disorder. \cite{fisher89}

At high temperatures (and in the thermodynamic limit $L \to \infty$, i.e., for
thick samples \cite{taeuber97}), one thus finds an entangled liquid of unbound
flux lines (corresponding to a boson superfluid), separated by a sharp
second-order transition from a low-temperature phase of localized vortices.
This {\em Bose glass} phase is characterized by an infinite tilt modulus
$c_{44}$, and turns out to be stable over a certain range of tipping angles of
${\bf B}$ away from the $z$ direction (transverse Meissner effect). 
\cite{nelson92} 
At low temperatures in the Bose glass, the dominant vortex transport mechanism
for low external currents is expected to be the analog of variable-range
hopping in disordered semiconductors, \cite{shklovskii84} namely the formation
of double kinks extending from one columar defect to another energetically
favorable, but not necessarily adjacent pinning site. \cite{nelson92}
Therefore the actual distribution of pinning energies constitutes an important
feature of the system that determines its low-energy current-voltage (I-V)
characteristics. 
As in the case of doped semiconductors, \cite{shklovskii84,davies82,moebius92}
long-range repulsive interactions drastically modify the distribution of
pinning energies, which is the analog of an (interacting) density of states,
and may lead to the emergence of a correlation-induced soft ``Coulomb gap''
separating the filled and empty states. \cite{taeuber95}
This effect in turn reduces vortex transport in the variable-range hopping
regime considerably.
Recent experiments have indeed identified the characteristic signatures of
variable-range hopping processes in irradiated superconductors. 
\cite{konczykowski95,vdbeek95,baumann96,thompson97}

\begin{figure}[t]
\centerline{\epsfysize 4cm \epsfbox{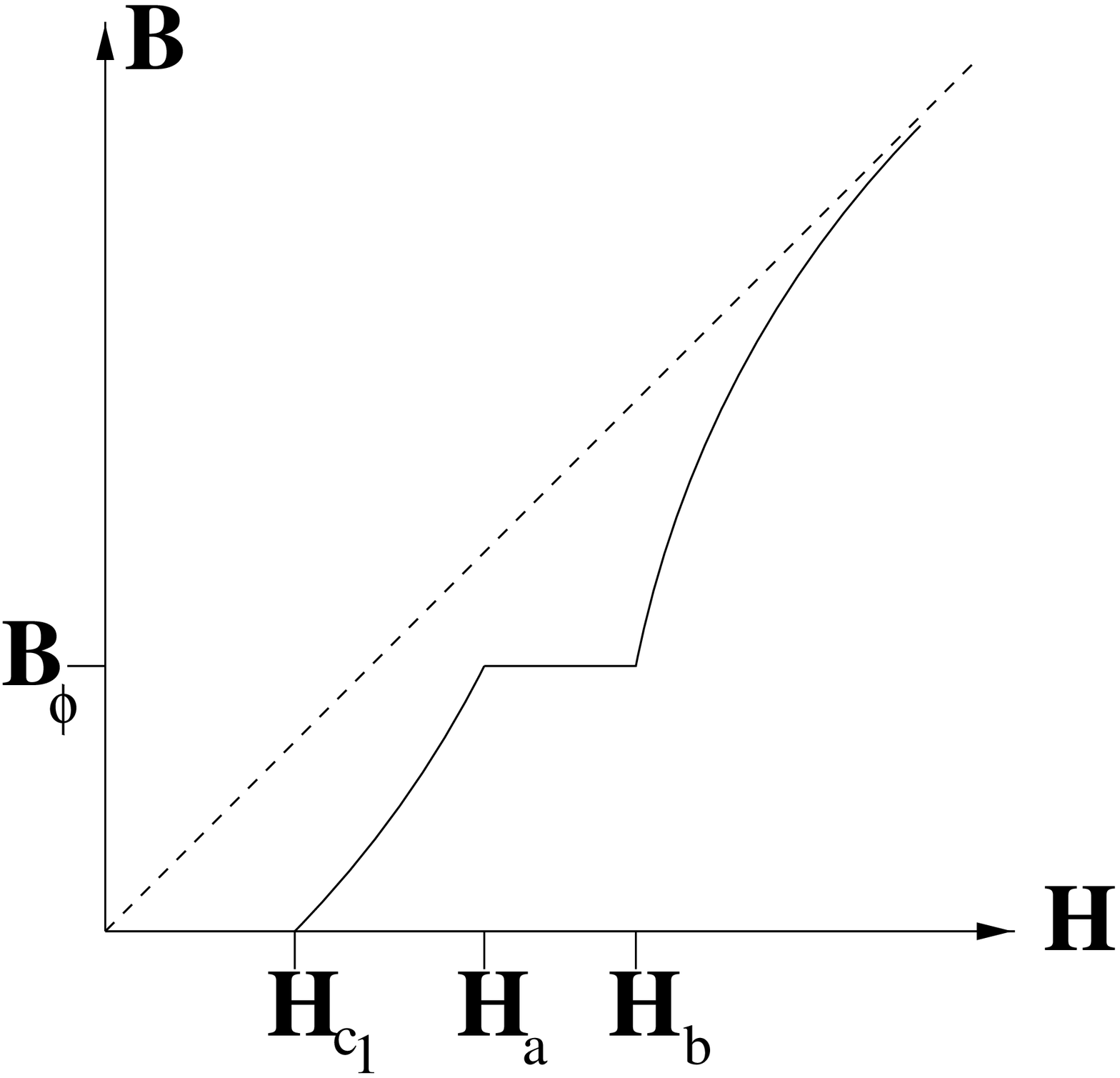} \qquad
            \epsfysize 4cm \epsfbox{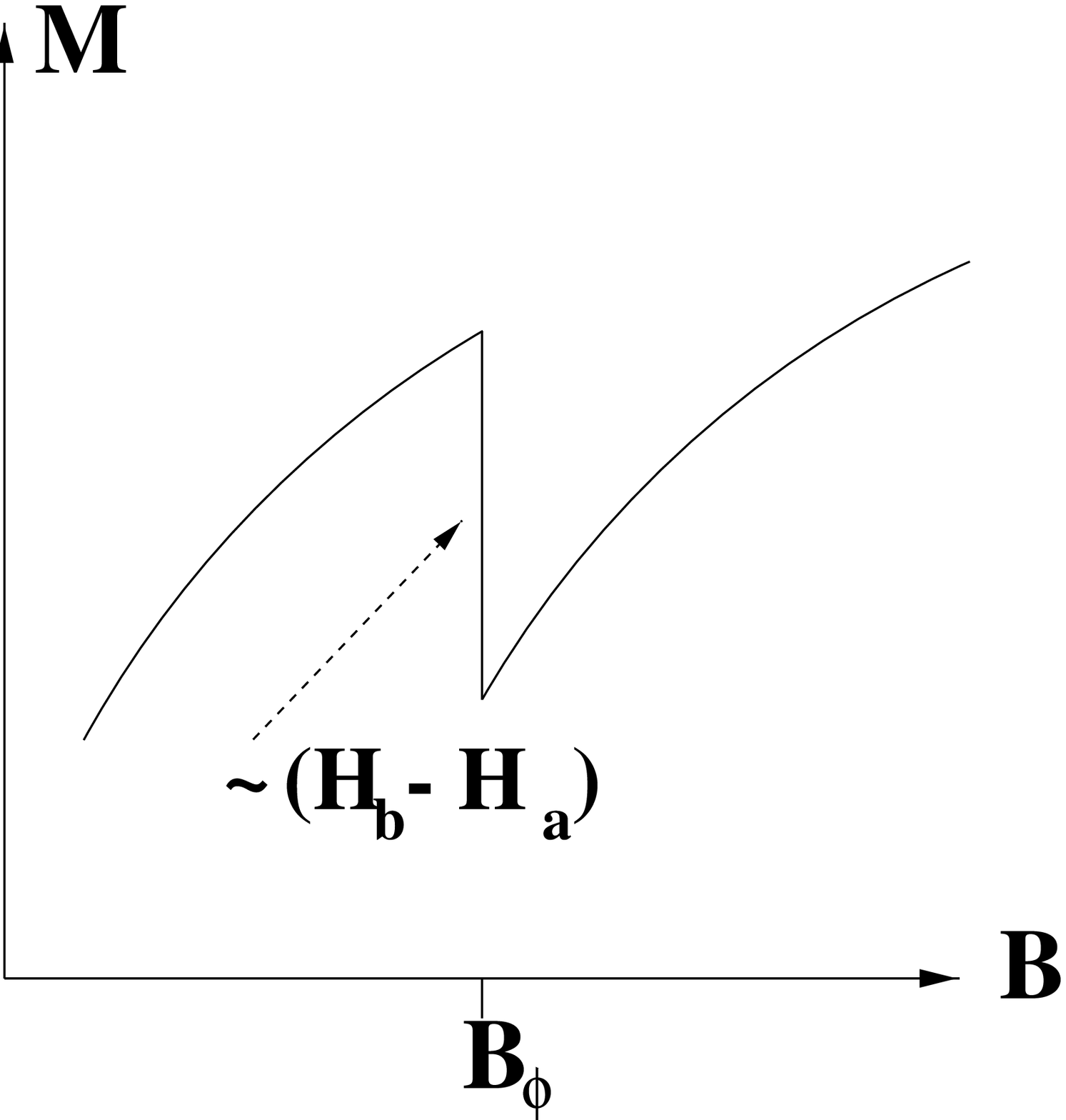}}
\caption{Left: $B$ vs $H$ showing the typical Mott insulator ``lock-in'' at the
  matching field $B_\Phi$. $B$ does not change over a certain range of the
  applied field $[H_a,H_b]$, until the magnetic field energy (at $H_b$) becomes
  large enough to push another vortex into the system. Right: At the Mott
  insulator transition, the magnetization $M = B - H$ should display a sharp
  jump of size $\propto (H_b - H_a)$ as a result of this ``lock-in'' effect.}
\label{mag_fig}
\end{figure}

At least for weakly interacting flux lines, the theory also suggests a
low-temperature {\em Mott insulator} phase, occurring when $B = B_\Phi$, i.e.,
the vortex density exactly matches that of the columnar damage tracks. 
\cite{nelson92}
In this situation, all the defects will be occupied by one vortex each, and the
introduction of another flux line into the system will require the typical 
pinning energy $U_0$ (provided the repulsive vortex interactions are
negligible --- weak repulsive forces will slightly increase this energy cost). 
Consequently, upon increasing the external field $H$, one expects a ``lock-in''
effect at $B = B_\Phi$, where the internal field remains constant until this
energy barrier is overcome, as shown in Fig.~\ref{mag_fig} (see also 
Refs.~[\onlinecite{wahl95,leo95,bulaevskii96,reichhardt96}]).
The Mott insulator phase is therefore characterized by a {\em hard gap} in the
distribution of pinning energies. \cite{nelson92}
Correspondingly, the (reversible) magnetization $M = B - H$ should display a
sharp downward jump {\em exactly} at $B = B_\Phi$; this can also be seen from 
the relation $M = - \partial G / \partial B$, where $G$ is the total Gibbs free
energy of the system, here dominated by the internal energy $E$, which 
increases abruptly at $B_\Phi$ as a new vortex enters the sample.
The Bose glass and the Mott insulator represent distinct thermodynamic phases,
for the latter should be characterized by an infinite tilt modulus $c_{44}$, 
{\em and} a diverging compressional modulus 
$c_{11} \sim \partial H / \partial B$. 
Characteristic minima in the reversible magnetization near the matching field
were indeed observed experimentally at low temperatures in single-crystal Tl
compounds, \cite{wahl95} and both BSCCO tapes \cite{li96} and single-crystals.
\cite{vdbeek96}
Furthermore, pronounced minima in the magnetization relaxation at 
$B \approx 1.4 B_\Phi$ were reported for YBCO single crystals,
\cite{beauchamp95} and for $B \approx 1.4 B_\Phi$ and $B \approx 3.3 B_\Phi$ in
a Tl compound, \cite{nowak96} and interpreted as further evidence for vortex
``lock-in'' behavior characteristic of the Mott insulator phase.

One of the main results of our numerical study will be, however, that the 
original predictions of a distinct thermodynamic Mott insulator phase are,
strictly speaking, applicable only in two limiting cases: 
(i) For short-range interactions, i.e., when the London penetration depth
    $\lambda$ is considerably smaller than the average defect spacing $d$, and
    the sole effect of the intervortex repulsion is to prevent double occupancy
    of columnar pins; but it cannot prevent all the defect sites, even for a
    spatially random distribution of pinning centers, to be filled with
    vortices. 
(ii) For a regular array of columnar defects, e.g., a triangular or square
    lattice, for which because of the spatial symmetry the interactions merely
    lead to a constant additive renormalization of the pinning energies,
    equal for all defect lattice sites.
In all intermediate cases, i.e., a disordered spatial defect distribution and a
sizeable interaction range $\lambda / d$, one has to expect that it will be
impossible to fill the columnar pins completely.
For, the occupation of very close pins will be energetically costly as compared
to moving a vortex into a high-symmetry ``interstitial'' position, at maximum
distance from neighboring defect sites.
Thus, a departure from both situations (i) and (ii) by either increasing the
interaction strength and range, or introducing disorder into the defect
configuration, will lead to a softening of the ideal Mott insulator hard gap in
the distribution of pinning energies, and smear the corresponding singularities
in the free energy $G$, magnetization $M$, and bulk modulus $c_{11}$.
In the thermodynamic sense, the Mott insulator does therefore not exist as a
distinct phase in typical experimental situations.
On the other hand, if $\lambda / d$ is not too large, we shall find that very
characteristic {\em remnants} of the Mott insulator singularities appear in
measurable quantities, and we shall indeed be able to explain why and in which
direction these maxima or minima are shifted away from $B = B_\Phi$, as
experimentally observed. \cite{wengel97}
Also, recently regular arrays of columnar defects have been produced in
high-$T_c$ materials, \cite{baert95} and in these more artificial samples the
Mott insulator phase might play a much more prominent role.

Another important effect of the vortex interactions leads to the distinction of
a strongly and a weakly pinned regime, respectively, within the Bose glass
phase. \cite{nelson92,leo95,larkin95}
For short-range interactions $\lambda / d \ll 1$, or for larger $\lambda / d$,
but low filling $f = B / B_\Phi \lesssim 0.5$, the vortices will be strongly
bound to the columnar defects, in the latter situation because of the
collective ``Coulomb gap'' correlation effects which render the transfer of a
flux line to another equally favorable site extremely unlikely. 
\cite{taeuber95} 
In the other extreme case, namely at large fillings $f > 1$ and for very strong
interactions, the pinning potentials actually constitute a rather weak
perturbation of the strongly correlated vortex system (with the flux lines
arranging themselves as closely as possible to a triangular lattice), and the
entire system can be shifted easily through the application of an external
current.
However, there should also be an intermediate weakly pinned Bose glass regime
around the matching field, where some of the flux lines are bound to the strong
pinning sites, and the others are held in place through their mutual repulsive
interactions with adjacent vortices. \cite{leo95,larkin95}
In a sample with a spatially random distribution of columnar defects at low 
temperatures, one would therefore expect a crossover from the Mott insulator to
a weakly pinned Bose glass, as $\lambda / d$ is increased.
Estimates for the location of this crossover line (denoted $B^*(T)$ there) are
given in Ref.~[\onlinecite{nelson92}].

Thus, if one aims to design a high-$T_c$ superconducting sample in such a way
that its pinning properties are optimized, clearly one needs to understand the
influence of the repulsive interactions and the effect of different spatial
pinning site distributions in a quantitative manner.
Obviously, there are two promising regimes for such a specific material
``tailoring'', namely either the underfilled ($f \lesssim 0.5$) strongly pinned
Bose glass where the ensuing soft correlation gap in the distribution of
pinning energies ensures that at least in the variable-range hopping regime
vortex transport is effectively reduced, or the incompressible Mott insulator
phase with its hard energy barrier preventing flux line motion.
It is the purpose of this paper, therefore, to provide a quantitative numerical
study of the low-temperature structural and transport properties of the entire
Bose glass phase, including and specifically addressing the region near the
matching field $B = B_\Phi$, as well as the features of the intermediate weakly
pinned regime.
Our investigations are based on a variant of the Coulomb glass model, which we
modify in such a way that high-symmetry interstitial sites are accessible to
the vortices in addition to the defect positions themselves, and we may thus
utilize the successful algorithms and considerable experience obtained in
earlier studies. \cite{shklovskii84,davies82,moebius92,taeuber95}
This model will be described in detail in Sec.~\ref{model-bg_sec}, while in
Sec.~\ref{sim-bg_sec} we shall introduce and define the quantities of interest
to characterize both relevant static and dynamic properties of the Bose glass
and Mott insulator phases in irradiated superconductors.
Sec.~\ref{databg_sec} is then devoted to the analysis of our simulation data
and contains the main results of our numerical study.
We shall explore both a random distribution of pinning sites (uniform disorder)
and spatially correlated disorder distributions, as obtained by deformations of
a perfect triangular defect lattice.
Finally, we shall summarize and discuss our results.

\section{The Model} \label{model-bg_sec}

The theoretical modeling of the Bose glass phase is based on the following
free energy for $N_V$ flux lines, described by their two-dimensional
trajectories ${\bf r}_i(z)$ as they traverse the sample of thickness $L$,
interacting with each other and with $N_D$ columnar defects parallel to the
magnetic field ${\bf B} \parallel \hat{{\bf z}}$, \cite{lyuksyutov92,nelson92}
\begin{eqnarray} \label{free_energy}
{\cal F} & = & \int_0^L \! dz \sum_{i=1}^{N_V} 
\left \{ \frac{\tilde{\epsilon}_1}{2} \left( \frac{d{\bf r}_i (z)}
{dz}\right)^{\! 2} + \frac{1}{2} \sum_{j\neq i}^{N_V} V[r_{ij}(z)]
\right. \nonumber \\ & &\qquad \qquad \qquad + \left. 
\sum_{k=1}^{N_D} V_D[{\bf r}_i (z) - {\bf R}_k] \right \} \ .
\end{eqnarray}
The first term in Eq.~(\ref{free_energy}) describes the elastic flux line 
tension and originates in an expansion of the line energy of nearly straight 
vortices with respect to small tipping angles. \cite{nelson89} 
The tilt modulus is given by $\tilde{\epsilon}_1 \sim \epsilon_0 \ln(a_0/\xi)$,
where the energy scale is set by $\epsilon_0 = (\phi_0/4\pi\lambda)^2$,
$a_0 = (4/3)^{1/4}(\phi_0/B)$ is the average lattice spacing between the flux
lines, $\xi$ is the superconducting correlation length, and $\lambda$ denotes
the London penetration depth.

The second term represents the interaction of all vortex pairs (which we
approximate to be local in $z$), where $r_{ij} = |{\bf r}_i - {\bf r}_j|$ and
$V(r) = 2\epsilon_0 K_0(r/\lambda)$ is the screened repulsive vortex potential,
with the modified Bessel function $K_0(x) \sim -\ln(x)$ as $x \to 0$, and
$K_0(x) \sim x^{-1/2}\exp(-x)$ for $x \to \infty$.

Finally, the $N_D$ columnar pins $\parallel {\bf B}$ are modeled by a
sum of $z$-independent attractive potential wells 
$V_D({\bf r} - {\bf R}_k) = - U_k \; \Theta(|{\bf r} - {\bf R}_k|)$, with
average spacing $d$, and centered at randomly distributed positions 
$\{{\bf R}_k \}$. 
Here, $\Theta$ denotes the Heaviside step function. 
The damage track diameters, produced by heavy ion irradiation, are typically
$2c_0\approx 100$ {\AA} wide, with a variation induced by the root-mean-square
energy dispersion of the ion beam of $\delta c_k/c_0\approx 15\%$, where $c_k$ 
denotes the individual defect radii. 
The pinning energies $U_k$ are related to the column diameters via the
interpolation formula given by Nelson and Vinokur \cite{nelson92}
\begin{equation}\label{intpol_eq}
U_k\approx \frac{\epsilon_0}{2} \ln[1 + (c_k/\sqrt{2}\xi)^2] \ ,
\end{equation}
where $\xi \approx 10$ {\AA} is the coherence length.
The distribution of column diameters thus induces a variation of the pinning
energies with the width 
\begin{equation}
w = \sqrt{\langle (\delta U_k)^2 \rangle } = 
\frac{\epsilon_0}{1 + (\sqrt{2}\xi/c_0)^2} \frac{\delta c_k}{c_0} \ .
\end{equation}
Above a certain temperature $T_0$, the effective radius of the
normal-conducting region at the defects will be given by the coherence length,
which in mean-field theory grows with temperature according to
\begin{equation}
\xi(T) = \xi_0 (1- T/T_c)^{-1/2} \ ,
\end{equation}
Hence, $\sqrt2 \xi(T)$ should be inserted into Eq.~(\ref{intpol_eq}) instead of
$c_k$, whenever $\sqrt2 \xi(T)\ge c_0$, or $T\ge T_0$, with 
\begin{equation}
T_0 = T_c (1-2\xi_0^2/c_0^2) \ .
\end{equation}
Consequently, above $T_0$ the fluctuations of the pinning energies will be 
effectively smoothed out.
With $\xi_0\approx10$ {\AA} and $c_0\approx50$ {\AA} this will happen at
$T_0\approx 0.92 T_c$. 
As we shall describe later, we will not be particularly interested in this
regime, since the Mott insulator is predicted to exist deep in the Bose glass
phase at low temperatures when thermal fluctuations can be neglected.

The classical statistical mechanics properties of Eq.~(\ref{free_energy}) can
be derived from an analysis of the grand-canonical partition function
\cite{nelson92} (we set the Boltzmann constant $k_{\rm B} \equiv 1$ throughout 
this paper)
\begin{eqnarray}
Z & = & \sum_{N_V=0}^\infty \frac{1}{N_V!} \exp(\mu {N_V}/T) \times
\nonumber \\
& & \int \prod_{i=1}^{N_V} {\rm d} {\bf r}_i(z) 
\exp\{-{\cal F}[{\bf r}_i(z)]/T\} \ , 
\end{eqnarray}
where $\mu$ denotes the chemical potential
\begin{equation}
\mu=\frac{\phi_0}{4\pi}H - \epsilon_0 \ln\kappa = 
\frac{\phi_0}{4\pi}(H-H_{c_1}) \ ,
\end{equation}
which changes sign at the lower critical field 
$H_{c_1} = \phi_0 \ln\kappa/(4\pi\lambda^2)$. 
The analytic approach to this partition function has made use of a mapping onto
a two-dimensional quantum-mechanical problem with a static (i.e., 
$z$-independent) disorder potential, where the physical direction $z$ has been
interpreted as imaginary time. \cite{lyuksyutov92,nelson92}  
In the thermodynamic limit $L\to \infty$ (which tunes the effective temperature
of the Bose system to zero), the properties of the model are determined by the
lowest-energy eigenstate of the corresponding transfer matrix or equivalent
quantum-mechanical problem. 
The corresponding ground state wave function is symmetric with respect to
exchange of flux lines, and thus of bosonic character. 
The ensuing localization transition found with this method, where the bosons
become bound to the static disorder potential (i.e., the flux lines are pinned
to the columnar defects), leads to a disorder-dominated low-temperature phase,
termed Bose glass.

Here, we will be interested in the low-temperature (ground state, $T=0$) 
properties of Eq.~(\ref{free_energy}), which is a fair approximation to
the Bose glass phase for temperatures $T \lesssim T_1$. 
$T_1$ denotes the temperature above which entropic corrections associated with
thermal fluctuations become relevant for the pinned flux lines. 
Physically, this means that vortices start to fluctuate around their defect
positions and can no longer be thought of as straight lines.
Theoretical estimates yield $T_1 \approx 0.6 \ldots 0.8 T_c$,
\cite{nelson92,taeuber95} although in some experiments $T_1$ has been found to
be considerably lower. \cite{krusin96,baumann96}
At an even higher depinning temperature $T_1<T_{\rm dp}<T_c$, vortices start to
leave the columnar defects, but may still be effectively pinned by their mutual
interactions with the neighboring occupied defects.
We restrict our study to the regime below $T_1$, where the vortices are
essentially straight lines, which simplifies our problem considerably.
When comparing our simulation results later with experimental data, 
$T \lesssim 0.6 \ldots 0.8 T_c$ will {\it a posteriori} prove to be a good
approximation. 

Since thermal wandering and bending of vortex lines below $T_1$ is suppressed,
we can neglect the bending energy term in Eq.~(\ref{free_energy}). 
This leaves us with a time-independent, two-dimensional problem of $N_V$
interacting ``particles'' and $N_D$ defects. 
We also aim to address a possible Mott insulator phase within the Bose glass,
near the matching field $B_\Phi$. 
Consequently, we want to study a filling fraction 
\begin{equation}
f=\frac{N_V}{N_D}=\frac{B}{B_\Phi} \approx 1 \ ,
\end{equation}
which is why we have to account for the possibility (especially for $f>1$) that
not all vortices can be accommodated on defect sites. 
Hence, we represent our problem on an underlying triangular grid with $N$
``lattice'' sites and $N-N_D$ interdefect positions (``interstitials''). 
This is the main new aspect of our simulations in comparison to those reported
in Ref.~[\onlinecite{taeuber95}], where only defect sites were available. 
Thus, that study was limited to low filling fractions $f \ll 1$. 
The effective, two-dimensional Hamiltonian used in our simulations, therefore
reads
\begin{equation} \label{heff_eq}
{\cal H}_{\rm eff} = \frac{1}{2} \sum_{i\neq j}^N n_i n_j V(r_{ij}) +
\sum_{k=1}^{N_D} n_k t_k \ ,
\end{equation}
and its grand-canonical counterpart is
\begin{equation} \label{heffmu_eq}
\tilde{\cal H}_{\rm eff} = {\cal H}_{\rm eff} - \mu \sum_{i=1}^{N} n_i \ .
\end{equation}
Here, $n_i= \{ 0,1 \}$ represents the site occupations number; note that 
$\sum_{i=1}^{N} n_i=N_V$. 
Taking $\xi\approx 10$ {\AA} (at $T=0$) and $c_0\approx 50$ {\AA}, we obtain
for the bare pinning energies $t_k=-\langle U_k\rangle + w_k$, with 
$\langle U_k\rangle =0.65$ and width $w=0.1$, both in units of $2\epsilon_0$. 
For simplicity, we assume a flat distribution of pinning energies around
its mean, $P(w_k)=\Theta(w-|w_k|)/(2w)$. 
Our choice of an underlying triangular grid is motivated by the fact that
without disorder the energetically most favorable state for the vortices is the
(triangular) Abrikosov lattice.
Thus in effect we take into account the energetically lowest high-symmetry
``interstitial'' positions between the columnar defects.

Finding the ground state configuration of this system by analytical
means constitutes a very difficult task, which has to date not been
successfully mastered. 
Consequently, we have to resort to numerical simulation techniques with a
suitable ground state finding algorithm. 
Fortunately, the Hamiltonian (\ref{heff_eq}) is precisely of the form studied
in the context of charge carriers localized at random impurities in doped
semiconductors (i.e., the so-called Coulomb glass problem with $V(r) \sim 1/r$
instead of our logarithmic repulsion), and we can utilize the broad numerical
experience gathered in those previous investigations.
\cite{shklovskii84,davies82,moebius92,taeuber95}
We shall use Eq.~(\ref{heff_eq}) in all of our simulations, and will now
introduce the simulation technique and the quantities we intend to measure.

\section{Simulation Technique and Quantities of Interest} \label{sim-bg_sec}

In this section we shall first describe in detail the Monte Carlo algorithm we
have employed in all simulations presented in this paper. 
We then proceed with an introduction to the relevant quantities we have 
measured, and which can be qualitatively and semi-quantitatively compared with 
experimental data that aimed to identify a Mott insulator transition in YBCO, 
BSCCO, and Tl-compounds.

\subsection{Energy Minimization Procedure}

Our starting Hamiltonian is given in Eq.~(\ref{heff_eq}). 
We will be interested in the ground state properties of this Hamiltonian as a
function of the filling fraction $f$, which we shall vary near the predicted
Mott insulator value $f=1$, and as a function of the interaction length scale
$\lambda$ or, more often, as function of the dimensionless ratio $\lambda/d$
where $d$ is the average defect distance.
In order to find the ground state configuration of this Hamiltonian, we follow
closely an energy minimization scheme described in detail by Shklovskii and
Efros in Ref.~[\onlinecite{shklovskii84}], Chap.~14, with some slight
modifications according to Davies, Lee, and Rice. \cite{davies82} 
Taking a triangular lattice with lattice constant $a \equiv 1$ and $N$ sites,
and imposing periodic boundary conditions, we initially distribute $N_D$
defects randomly on this underlying triangular grid. 
Each defect site is assigned a randomly chosen energy $t_k$, from a flat
distribution with mean $-\langle U_k \rangle /(2\epsilon_0) = -0.65$ and width
$w/(2\epsilon_0)=0.1$. 
Most of our simulations used $N=1600$ and $N_D=100$, such that the average
defect distance was $d=4$, in units of the lattice constant $a \equiv 1$.

We then distribute $N_V=fN_D$ vortices randomly on the grid and follow the
algorithm by Shklovskii and Efros in order to minimize the energy. 
This is done as follows: 
First, for our interacting system, we define single-particle site energies as 
\begin{equation} \label{locen_eq}
\epsilon_i = \frac{\partial {\cal H}_{\rm eff}}{\partial n_i} =
\sum_{j\neq i}^N n_j V(r_{ij}) + t_i \ .
\end{equation}
For a filled site, $\epsilon_i$ denotes the energy to remove a particle from 
site $i$ to infinity, while, for an empty site, $\epsilon_i$ is the energy 
needed to introduce an additional particle from infinity on site $i$. 
In thermal equilibrium, the chemical potential $\mu$ energetically separates
the occupied from the unoccupied states, i.e., $\epsilon_i\le \mu \forall i$
with $n_i=1$ and $\epsilon_i\ge\mu \forall i$ with $n_i=0$. 
In the random initial configuration, which usually corresponds to a strongly
non-equilibrium situation, many occupied sites will have a higher energy then
many unoccupied sites.
To approximate the ground state as closely as possible, we proceed in two 
steps. 
After the calculation of all $\epsilon_i$ we try to relax the initial state by
successive particle-hole transitions. 
To do so, we determine the occupied site $p$ with highest energy,
$\epsilon_p=\max_{\{n_i=1\}}\epsilon_i$ and the empty site $q$ with lowest
energy, $\epsilon_q=\min_{\{n_i=0\}}\epsilon_i$. 
If $\epsilon_q < \epsilon_p$, the particle on site $p$ is moved to site $q$,
the site energies are recalculated and the procedure is repeated until
eventually 
$\epsilon_p\le\epsilon_q\,\forall p\in\{i|n_i=1 \}\;,\forall q\in\{i|n_i=0\}$. 
Thus we have now separated the occupied from the unoccupied states
energetically and can give a first estimate of the chemical potential by
\begin{equation} \label{mu_eq}
\mu=\frac{\epsilon_q + \epsilon_p}{2} \ .
\end{equation}

The ensuing state is, however, still a very poor approximation of the ground 
state, as it is generally unstable against single-particle hops.  
To test this, we next compute the hopping energies 
\begin{equation} \label{hop_eq}
\Delta_{i\to j} = \epsilon_j - \epsilon_i - V(r_{ij}) \ .
\end{equation}
This expression can be understood as follows. 
First, remove a particle from site $i$ to infinity, which yields an energy gain
of $\epsilon_i$. 
Then, take the particle back to site $j$, which costs the amount of 
$\epsilon_j-V(r_{ij})$. 
The additional contribution of $V(r_{ij})$ stems from the fact that after
removing the particle from site $i$ there are only $N_V-1$ particles left, but
$\epsilon_j$ is defined in terms of a system with $N_V$ particles. 
Thus, the fictitious interaction with site $i$ has to be accounted for
explicitly. 
Note that a {\em necessary condition} for the ground state is 
$\Delta_{i\to j} > 0$ for all pairs $(i, j)$ with $n_i=1, n_j=0$.

\begin{figure}[t]
\centerline{\epsfxsize 4.6cm \epsfbox{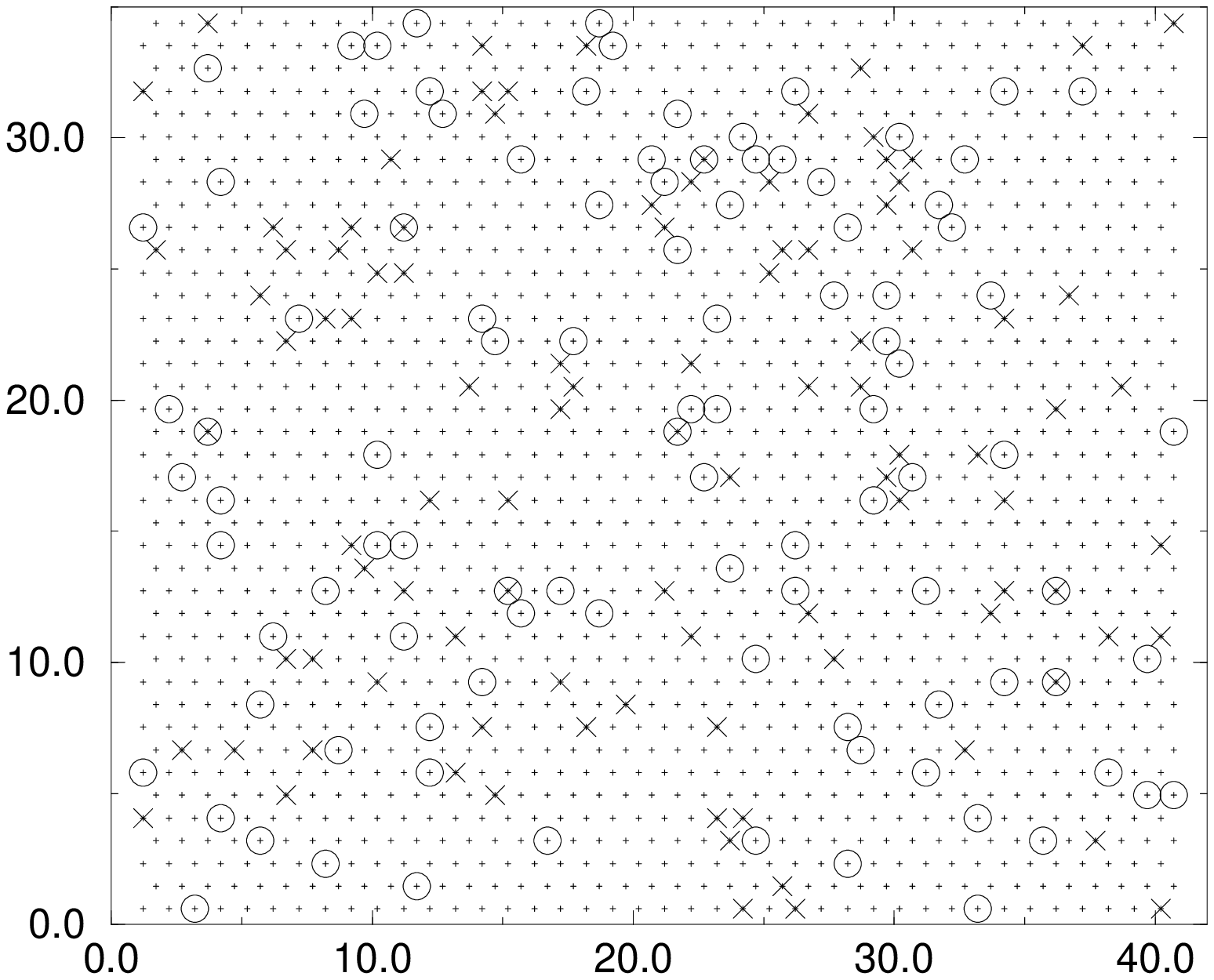}
            \epsfxsize 4.6cm \epsfbox{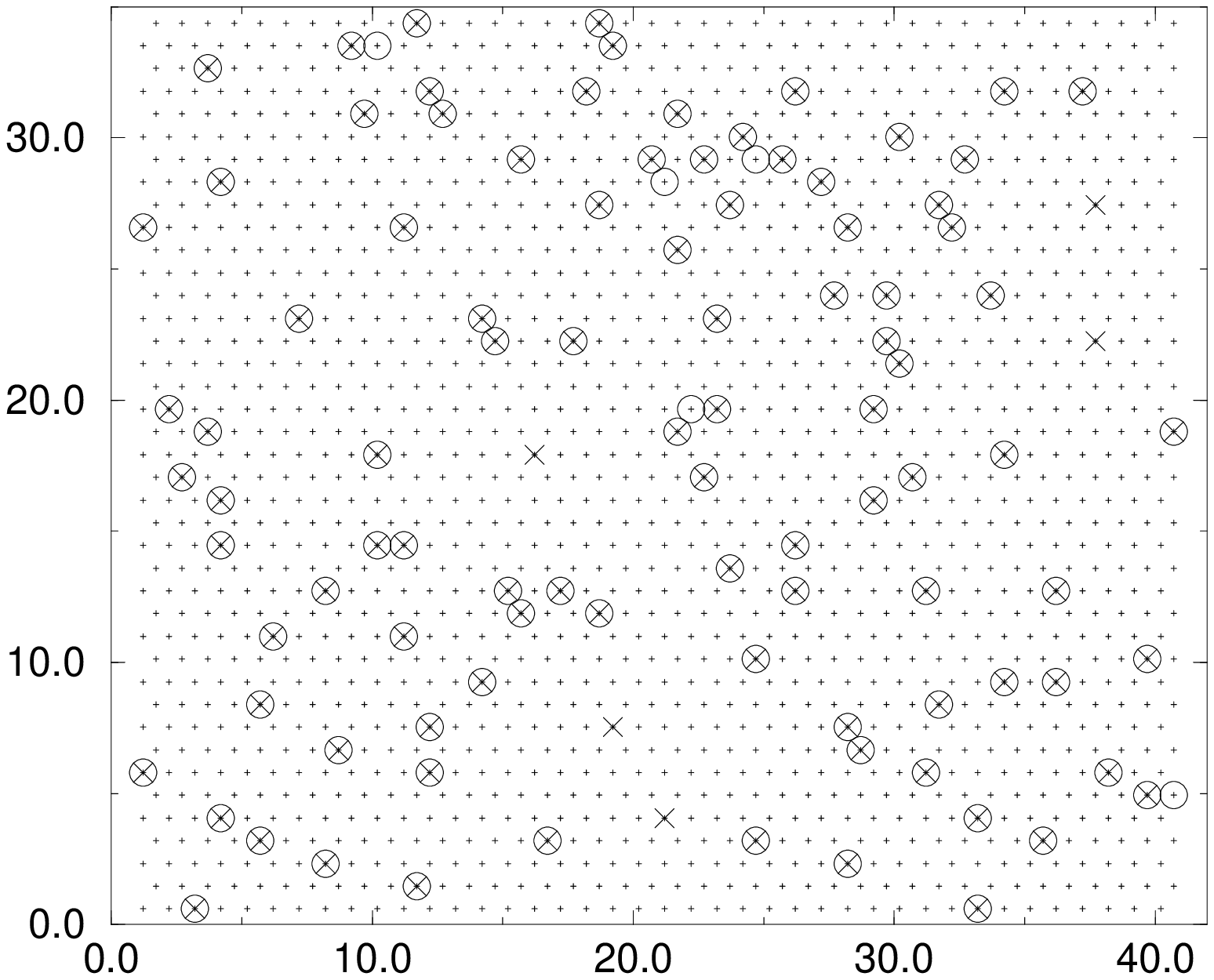}}
\caption{Left: Typical initial configuration of randomly distributed defects
  ($\circ$) and vortices ($\times$) on a triangular lattice (dots) with 
  $N=1600$ and $N_V=N_D=100$, i.e., $f=1$; here we set $\lambda/d=1$. 
  Right: Final configuration after the relaxation procedure. Most of the
         vortices are now located on energetically favorable defect sites,
         except for five particles, since the respective defect sites are too
         close together, such that one of two/three close vortices is pushed
         away into low-energy ``interstitial'' voids due to the repulsive 
         interaction $V(r_{ij})$.} 
\label{config_pic}
\end{figure}

Thus, in a second step we compute all $\Delta_{i\to j}$, and then search for
the minimum of all the negative hopping energies: 
$\Delta_{p\to q}=\min_{\{\Delta_{i\to j}<0\}} \Delta_{i\to j}$. 
Then, we perform the particle transfer from site $p$ to site $q$, which reduces
the total energy $E$ of the system further. 
As this hop typically leads to a non-equilibrium state again, we next
recalculate the site energies and repeat the particle-hole transfer procedure
of the preceding paragraph.
Then, another hopping transfer is attempted until eventually 
$\Delta_{i\to j}>0\,\forall i\neq j$. 
The ensuing configuration is accepted as a fair approximation of the ground
state for a particular defect distribution, filling fraction $f$ and ratio
$\lambda/d$. 
A typical initial and final particle distribution obtained with this procedure
is shown in Fig.~\ref{config_pic}.
From the resulting configuration the site energies $\epsilon_i$, the chemical
potential $\mu$, and the total energy $E$ of the system can be computed.
This procedure is repeated for a number of $k$ different initial realizations
of the disorder, such that we can obtain an ensemble average of the chemical
potential $[\mu], [E]$ and other quantities of interest, which we shall 
describe in the following subsections. 
We typically take $k=50$ which has proven to be sufficient in our simulations.

Of course, with this approximation scheme we still may not have reached the
true ground state configuration, but rather a metastable ``pseudo''-ground
state. 
In principle, one would have to test that pseudo-ground state against
$m=2,3,...N_V$ particle hops, in order to obtain the true minimum-energy
configuration.
Previous investigations have shown, though, that terminating at $m=1$ already
gives reliable estimates for the energy-level distribution. 
\cite{shklovskii84,davies82,moebius92} 
To test the algorithm, we have equilibrated different initial configurations of
vortices with an identical defect distribution. 
For example, we chose initially a) a completely random configuration, b) a 
triangular array, and c) we first occupied all defects with vortices, and then
distributed the remaining vortices (if $f>1$) randomly, for given $f$ and 
$\lambda/d$. 
All different configurations yielded the same energies and $\mu$ within the
numerical uncertainties, although the spatial configurations typically looked
slightly different.  
Additional tests of the algorithm will be presented in Sec.~\ref{databg_sec}. 

Having introduced our simulation algorithm, we will employ this technique for
$0.3\le f\le 3$ and $\lambda/d=1/4,1/2,1,5$ \cite{footnote1} in order to 
investigate whether there exists a Mott insulator within the Bose glass phase, 
and see if it is stable against increasing the vortex-vortex repulsion as 
$\lambda/d$ is enhanced, while keeping the pinning strengths $U_k$ fixed.
In order to identify the Mott insulator and to compare with experimental data,
we need to introduce a number of quantities that we can compute from our ground
state configurations. 
This will be the topic of the next four subsections.

\subsection{Distribution of (Interacting) Pinning Energies}

From the interacting single-site energies defined in Eq.~(\ref{locen_eq}) we
can construct a histogram of all energies of the final ground state
configuration. 
The quantity $g(\epsilon){\rm d} \epsilon$ then denotes the number of states
per unit area with energies in the interval 
$[\epsilon,\epsilon+{\rm d}\epsilon]$ with the normalizing constraint  
\begin{equation}
\int_{-\infty}^{\infty} g(\epsilon) {\rm d}\epsilon = 1/a^2 \equiv 1 \ .
\end{equation}
The quantity $g(\epsilon)$ thus defines the interacting boson single-particle
density of states or, in the flux line context, the distribution of interacting
pinning energies.
This quantity is particularly interesting in order to identify a Mott 
insulator. 
For a true Mott insulator phase, $g(\epsilon)$ should display a hard gap
$\Delta$ near the chemical potential $\mu$, i.e., the (lower) energies of the
occupied states should be separated from the (higher) energies of the
unoccupied ``band'' by a region of finite width which contains no states at
all. 
In the strong screening limit $\lambda\to 0$, when there is no interaction
between the vortices, the gap should be exactly 
$\Delta = \langle U_k \rangle -w$ (see the top left of Fig.~\ref{dos_fig} 
below). 

%\pic{
%\centerline{\epsfxsize 10 cm \epsfbox{emu.eps}} 
%\mycap{Density of states for $f=0.8$ and $\lambda/d=1$.}
%\label{emu_fig}
%}

Moreover, by further exploiting the analogy to two-dimensional bosons localized
at randomly distributed defect sites, one has to consider the possibility that
these spatial correlations produce a soft Coulomb gap in the distribution of 
pinning energies near the chemical potential $\mu$. 
Such a Coulomb gap has indeed been found in the Coulomb glass problem of
disordered semiconductors, where the interaction is also long-range 
($\sim 1/r$ up to the screening length $\lambda$), 
\cite{shklovskii84,davies82,moebius92} and in the Bose glass phase for small 
fillings $f$ and large $\lambda/d$, i.e., almost logarithmic interactions. 
\cite{nelson92,taeuber95} 
In both situations the density of states vanishes precisely at the chemical
potential, and in its vicinity increases according to a power law, 
\begin{equation}
\label{softgap_eq}
g(\epsilon) \sim |\epsilon - \mu|^{s_{\rm eff}} \ .
\end{equation}
Here, we have defined the {\em effective gap exponent} $s_{\rm eff}$, which 
depends strongly on $\lambda$ and on the filling fraction $f$, as we shall see.
In fact, for small fillings $f < 1$ and short-range interactions 
$\lambda \to 0$, the exponent is trivially $s_{\rm eff}=0$, i.e., the density
of states near $\mu$ is constant. \cite{nelson92,taeuber95} 
For $\lambda\to\infty$, the theoretical prediction within the framework of a
mean-field type of analysis would be $s_{\rm eff}=\infty$. 
%In Fig.~\Ref{emu_fig} we plot a sample density of states for $f=0.8$ and
%$\lambda/d=1$. On the left ($\epsilon - \mu < 0$) are the occupied
%states and on the right ($\epsilon - \mu > 0$) are the unoccupied
%states, divided by a soft Coulomb gap near $\mu$. 
With a random distribution of pinning sites, however, one has to expect that
local energy fluctuations will play an important role, and thus 
$0 < s_{\rm eff} < \infty$, with values that may depend on the filling $f$, and
the actual energy scale. \cite{moebius92,taeuber95}

For finite $\lambda / d$, and sufficiently low energies in the utmost vicinity
of $\mu$, one must find a crossover to $s_{\rm eff}=0$ in an infinite system.
In a finite system, however, or at some distance from the chemical potential,
one will find a nonzero {\em effective} exponent $s_{\rm eff}$, which can be
obtained simply by fitting a power law in the vicinity of $\mu$ (see also
Ref.~[\onlinecite{taeuber95}]).
The determination of $s_{\rm eff}$ depending on $\lambda/d$ and $f$ is one of
the tasks of our work since it affects crucially the experimentally measurable
I-V characteristics of superconductors in the Bose glass phase, as we shall see
below. 
Primarily, though, we want to see whether a hard gap $\Delta$ in the density of
states survives the introduction of long-range interactions.

\subsection{Bulk Modulus}

From our measurement of the chemical potential $\mu$ as a function of filling
fraction $f$ and constant interaction scale $\lambda/d$, we can determine the
bulk modulus $c_{11}$ via 
\begin{equation}
\label{bulk_eq}
c_{11} = \frac{\partial \mu}{\partial f} = N_D \frac{\partial \mu} 
{\partial N_V} = \tilde{\kappa}^{-1} \ ,
\end{equation}
where $\tilde{\kappa}$ denotes the compressibility of the system.
Physically, the bulk modulus characterizes the stiffness of the system with
respect to the insertion of an additional particle into it. 
A divergence of $c_{11}$ at the matching field $B=B_\Phi$ ($f=1$) would 
indicate the presence of the Mott insulator phase, since the chemical potential
$\mu$ should show a sharp jump when one more vortex is put into the system. 
In the case of $\lambda\to 0$, the jump becomes of the size of the hard gap in
$g(\epsilon)$, $\Delta = \langle U_k \rangle -w$, i.e., introducing another
vortex into the system costs a finite energy $\Delta$. 
In this respect, the system becomes similar to a Meissner phase which has zero
compressibility as well.

\subsection{Reversible Magnetization}
\label{rmag_ssec}

Since near the Mott insulator the Bose glass should become more and more
diamagnetic due to the diverging bulk modulus, similar to the situation in the 
Meissner phase, a number of experiments have tried to identify the Mott 
insulator by measuring the magnetization as a function of the (internal) 
magnetic field $M(B)$. \cite{wahl95,li96,vdbeek96}
In our simulations we can also obtain the reversible magnetization as a
function of $f$ from the total free energy of the system $G$, which becomes 
identical to the internal energy $E$ at $T = 0$, by noting the thermodynamic 
relation  
\begin{equation}
\label{mag_eq}
M=-\frac{\partial G}{\partial B} \sim - \frac{1}{N_V}
\frac{\partial G}{\partial f} \ .
\end{equation}
Whereas in unirradiated samples the magnetization increase with $f$ should be
proportional to $\ln f$, \cite{tinkham75,vdbeek96,li96,bulaevskii96} the 
signature of a true Mott insulator would be a sharp negative dip in the 
magnetization at $f=1$ due to the diamagnetic effect (diverging bulk modulus), 
see Fig.~\ref{mag_fig} above.
Consequently, the bulk modulus and the magnetization are in fact closely
related to each other. 
Noting that $M = B - H$, $f \sim B$, and $\mu \sim H - H_{c_1}$, we can write
\begin{equation}
\label{bm-mag_eq}
c_{11} = \frac{\partial \mu}{\partial f} \sim \frac{\partial H}{\partial B} 
\sim 1 - \frac{\partial M}{\partial B} = 1 + \frac{\partial^2 G}{\partial B^2} 
\ . 
\end{equation}
Thus, the maximum of $c_{11}$ occurs at the location of the steepest negative
slope in $M(B)$, which usually is at a lower $B$ value than the minimum in $M$.
Only in the true Mott insulator phase will the singularities in $c_{11}$ and
$M$ coincide {\em precisely} at $B=B_\Phi$.

\subsection{I-V Characteristics and Magnetization Relaxation}
\label{iv+mag_ssec}

The I-V characteristics measured in the Bose glass phase below $T_c$ have the
same form as predicted for the vortex glass, namely 
\begin{equation}
\label{iv_eq}
E=\rho_0 J \exp[-U(J)/T] = \rho_0 J \exp[-(U_c/T)(J_1/J)^{p_{\rm eff}}] \ ,
\end{equation}
where $\rho_0$ is the normal state resistivity, $U_c$ is an energy barrier
scale, $J_1$ a current density scale and most importantly, $p_{\rm eff}$
denotes the {\em effective glass} or {\em transport exponent}. \cite{nelson92}
Experimentally, one determines the glass exponent in different filling regimes
and for different temperatures (effective interaction length scales $\lambda$).
We will see in this section that at low currents $p_{\rm eff}$ is closely
connected to $s_{\rm eff}$ (in the variable-range hopping regime, see below) 
via the (approximate) equation
\begin{equation} \label{peff}
p_{\rm eff} = \frac{s_{\rm eff} + 1}{D + s_{\rm eff} + 1} \ ,
\end{equation}
where $D$ is the dimension perpendicular to the applied field (here $D=2$).
In order to understand how we can extract the glass exponent from our data of
the density of states, we first have to distinguish between two current density
regimes.
The discussion presented here essentially follows 
Refs.~[\onlinecite{nelson92,taeuber95}].

\subsubsection{Hopping via vortex half-loop formation}

\begin{figure}[t]
\centerline{\epsfxsize 3.5cm\epsfbox{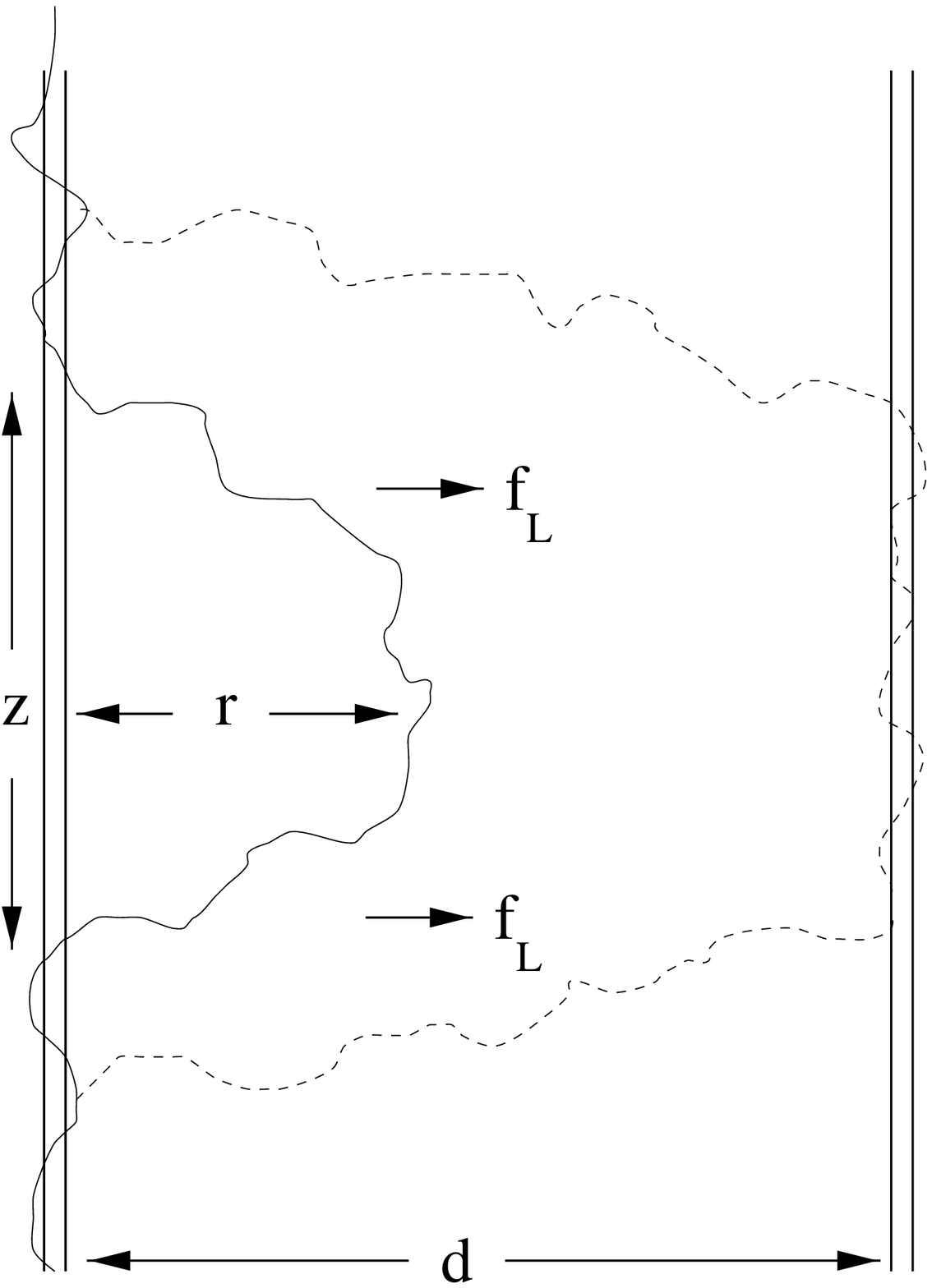}\qquad \qquad 
            \epsfxsize 3.5cm\epsfbox{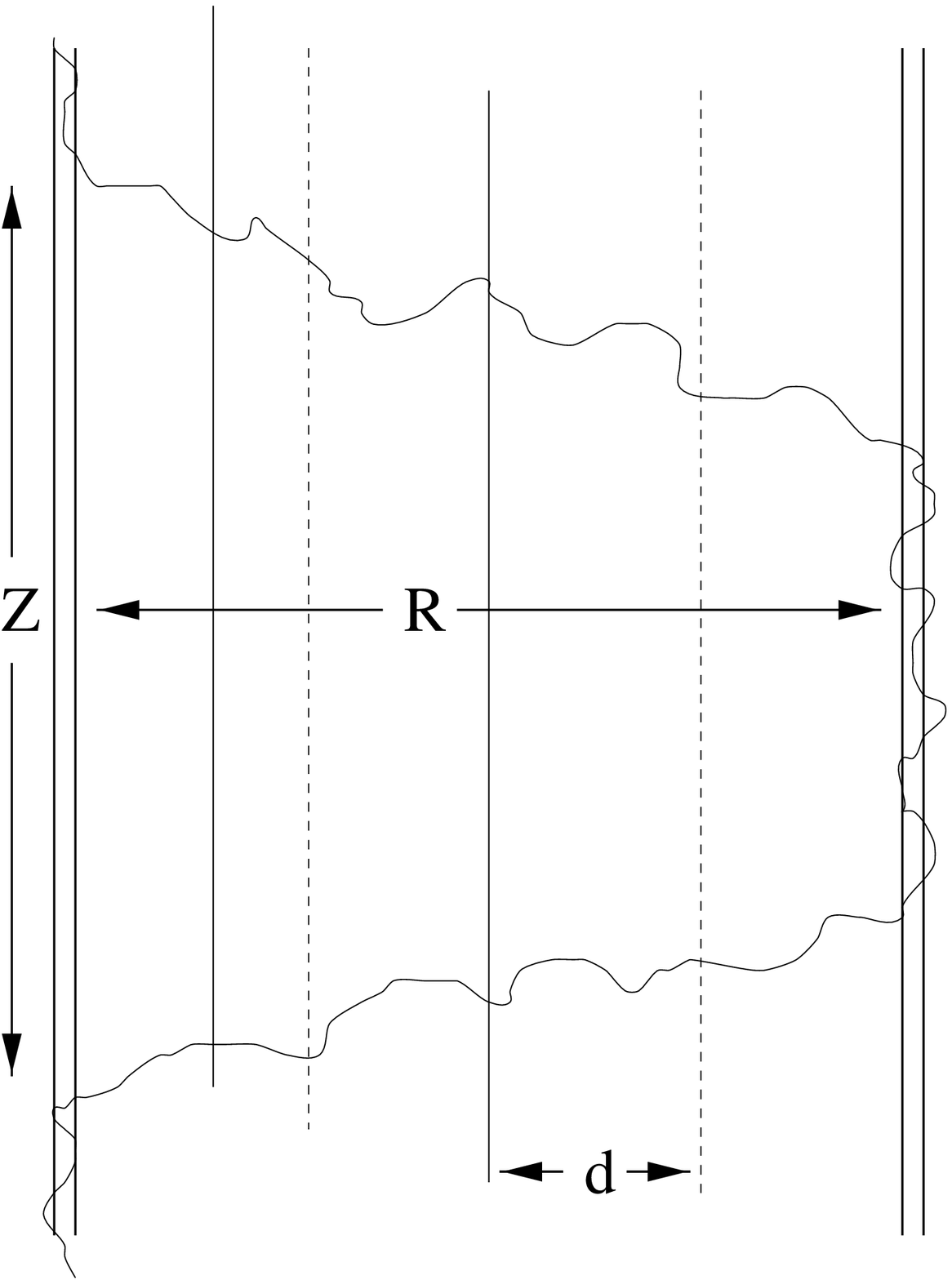}}
\caption{Left: Two columnar defects of distance $d$ with a vortex half-loop
  (solid line), and a double-kink excitation (dashed line), respectively. 
  The Lorentz force $f_L$ attempts to push the vortices to the right. 
  Right: Double-superkink configuration, as required for variable-range
  hopping. The flux line ``tongues'' seek out a compatible low-energy, but not 
  necessarily adjacent pin at distance $R$.} 
\label{kinks_fig}
\end{figure}

Let us first consider fairly strong currents $J_1<J<J_c$, where $J_c$ is the
critical depinning current above which superconductivity is destroyed. 
We shall see that in this regime, the motion of a single vortex is unaffected
by the other flux lines in the sample. \cite{nelson92}
A current perpendicular to the vortices ${\bf J}\perp {\bf B}$ will induce a
Lorentz force per unit length 
${\bf f}_L = (\phi_0/c)\hat{{\bf z}}\times {\bf J}$.
Thus we add to our model free energy of Eq.~(\ref{free_energy}) the term
\begin{equation} \label{lor-free_eq}
\delta {\cal F}_L = -\int_0^L {\rm d} z \sum_{i=1}^{N_V} {\bf f}_L\cdot
{\bf r}_i(z) \ .
\end{equation}
Let us now estimate the effective glass exponent from the saddle-point
configuration of Eq.~(\ref{free_energy}), supplemented with 
Eq.~(\ref{lor-free_eq}). 
A flux line will start to leave its columnar defect due to the current by
detaching a segment of length $z$ into the defect-free region, thereby forming
a half-loop of characteristic transverse size $r$, see 
Fig.~\ref{kinks_fig} (left). 
This costs the bending free energy $\tilde{\epsilon_1}r^2/z$, and the lost 
pinning energy $U_0 z$. 
This estimate uses on the average pinning potential $U_0=\langle U_k \rangle$.
Taking the work by the Lorentz force into account, the free energy of
the loop becomes approximately \cite{nelson92}
\begin{equation} \label{delfree_eq}
\delta{\cal F}_> (r,z) \approx \tilde{\epsilon_1}r^2/z + U_0 z - f_L rz \ .
\end{equation}
Optimizing this expression in the absence of external currents leads to the
saddle-point configuration $z^* \approx r^* \sqrt{\tilde{\epsilon}_1}/ U_0$,
while for finite currents one finds $r^*\approx cU_0/(J \phi_0)$, which after
inserting into Eq.~(\ref{delfree_eq}) yields
\begin{equation} \label{delfreeop_eq}
\delta {\cal F}^*_> \approx \frac{c \sqrt{\tilde{\epsilon}_1}U_0^{3/2}}
{J \phi_0} \ .
\end{equation}
We now identify the saddle-point free energy $\delta {\cal F}_>^*$ with the 
typical energy barrier $U(J)$. 
For a sufficiently low current $J_1$, the vortex half loop will extend to the
nearest-neighbor pin of distance $r^* = d$, and hence $J_1=c U_0/(\phi_0 d)$. 
The flux line will then form a double kink of width
$w_k=d\sqrt{\tilde{\epsilon_1}/U_0}$, which costs the free energy 
$2E_k=2d\sqrt{\tilde{\epsilon_1} U_0}$. 
Thus, for $J_1<J<J_c$, with these expressions and Eq.~(\ref{delfreeop_eq}), we
arrive at a current-voltage characteristics of the form (\ref{iv_eq}) with the
effective exponent $p_{\rm eff}=1$. 
Recent experiments on BSCCO samples have indeed found $p_{\rm eff}=1$ for
intermediate currents, supporting this half-loop nucleation picture.
\cite{konczykowski95,vdbeek95,baumann96} 

\subsubsection{Variable-range hopping of flux lines}

For $J<J_1$, the most important thermally activated excitations will be double
``superkinks'': The vortex will send out a tongue-like pair of kinks to a
possibly very distant neighbor, but with almost the same pinning energy; 
see Fig.~\ref{kinks_fig} (right). 
Once the tongue is established, the superkinks move outward and the vortex
hops to a new site.
This process represents the flux-line analog of {\em variable-range} charge
hopping transport between localized states in semiconductors.
\cite{shklovskii84} 
The free energy cost of a double superkink of transverse size $R$ and extension
$Z$ stems from the elastic term $2E_k(d \to R) = 2E_k R/d$, and the difference 
of pinning energies between the highest occupied state $\epsilon_i\approx \mu$ 
and the empty site energy $\epsilon_j\approx \mu + \Delta (R)$ at distance $R$.
In the presence of a small external current we can thus write the free energy
as \cite{nelson92}
\begin{equation} \label{del-free2_eq}
\delta {\cal F}_< \approx 2E_k R/d + Z \Delta (R) - f_L R Z \ .
\end{equation}
The density of available states as a function of $R$ with dimension $D$
transverse to ${\bf B}$ ($D=2$) on the one hand equals 
$d^D \int_{\mu}^{\mu+\Delta (R)} g(\epsilon)\, {\rm d}\epsilon$ from
considering the energetics, and on the other hand is given by $\approx (d/R)^D$
from simple statistics in real space. 
$\Delta(R)$ can then be determined from the equation
\begin{equation} \label{den-int_eq}
\int_{\mu}^{\mu+\Delta (R)} g(\epsilon)\, {\rm d}\epsilon \approx R^{-D} \ .
\end{equation}

We now proceed by optimizing Eq.~(\ref{del-free2_eq}) for $J=0$ first, which
yields the optimal longitudinal extent
$Z^* \approx 2 E_k/[d\,  (\partial \Delta/\partial R)_{R^*}]$, while for a
non-zero current one arrives at
\begin{equation} \label{jcrit_eq}
J\phi_0/c \approx\Delta(R^*)/R^* \ ,
\end{equation}
which through inversion gives the typical hopping range $R^*(J)$. 
Inserting this last result back into Eq.~(\ref{del-free2_eq}), we find the
typical barrier for a jump, and consequently \cite{taeuber95}
\begin{equation} \label{iv-int_eq}
E = \rho_0 J \exp[- \delta {\cal F}_<^*/T]= \rho_0 J 
\exp[-(2 E_k/T) (R^*(J)/d)] \ . 
\end{equation}

This procedure can be followed numerically, starting with a given distribution
of pinning energies $g(\epsilon)$, as obtained from the energy-minimization
algorithm.
In order to determine the I-V exponent explicitly for a density of states of
the form $g(\epsilon) =\gamma |\epsilon - \mu|^{s_{\rm eff}}$, we insert this
into Eq.~(\ref{den-int_eq}) to get
\begin{equation}
  \Delta (R) = \left(\frac{s_{\rm eff}+1}
{\gamma}\right)^{1/(s_{\rm eff}+1)} R^{-D/(s_{\rm eff}+1)} \ ,
\end{equation}
and with Eq.~(\ref{jcrit_eq}) finally arrive at
\begin{equation} \label{rstar_eq}
R^*(J) = \left(\frac{s_{\rm eff}+1}{\gamma}\right)^{1/(D+s_{\rm eff}+1)}
\left(\frac{c}{\phi_0 J}\right)^{p_{\rm eff}} \ ,
\end{equation}
with $p_{\rm eff}$ given by Eq.~(\ref{peff}).
Thus, the I-V characteristics will mainly be determined by the
interaction-dependent effective exponent $p_{\rm eff}$, which itself depends on
the effective gap exponent $s_{\rm eff}$, i.e., the amount of depletion of the 
density of states in the vicinity of $\mu$. 
With Eq.~(\ref{iv-int_eq}) we can write the I-V characteristics finally in the
form 
\begin{equation} \label{ivfinal_eq}
E = \rho_0 J \exp[-(2E_k/T) (J_0/J)^{p_{\rm eff}}] \ ,
\end{equation}
with $J_0= (c/\phi_0)\, \gamma^{1/(s_{\rm eff}+1)}\, d^{1/p_{\rm eff}}$.
For the non-interacting case (with a constant density of states, i.e., 
$s_{\rm eff}=0$), which is experimentally realized for very low magnetic fields
$f \approx 0.01$, and thus $\lambda/a_0\to 0$, one obtains the {\em Mott
variable-range hopping exponent} $p_{\rm eff}=1/3$.
Thus, we expect $p_{\rm eff}$ to vary in the interval $[1/3,1]$ (the upper
limit being given by $s_{\rm eff} \to \infty$), depending on the interaction
range $\lambda$, {\em and} the filling fraction.  
This has indeed been found in recent experiments, \cite{baumann96,thompson97} 
and partly in simulations. \cite{taeuber95} 

\begin{figure}[t]
\centerline{\epsfxsize 7.5cm \epsfbox{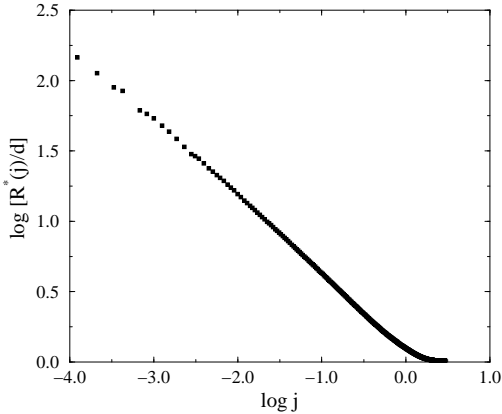}}
\caption{Double-logarithmic plot (base 10) of the exponential factor 
  $R^*(j)/d \propto \delta F_>^*(j)$ vs the reduced current $j=J\phi_0d/c$,
  derived numerically from the site-energy distribution $g(\epsilon)$.
  Here, $\lambda / d = 1$, and $f = 0.8$.}
\label{iv_fig}
\end{figure}

With the distribution of site energies as obtained from our minimizing 
procedure, we can numerically evaluate the I-V characteristics for any form of 
$g(\epsilon)$. 
This is done by integrating over $g(\epsilon)$ starting at $\mu$ and using the
relations (\ref{den-int_eq}) -- (\ref{ivfinal_eq}) to obtain $R^*(J)$.
A typical result derived from the distribution of pinning energies is depicted
in Fig.~\ref{iv_fig}, which shows a double-logarithmic plot of the effective 
hop size $R^*(j)/d \sim \delta F_>^*(j) \sim U(j)$ as a function of the 
reduced current $j = J\phi_0d/c$. 
According to Eq.~(\ref{rstar_eq}), the slope, determined by linear regression,
immediately yields the glass exponent $p_{\rm eff}$. 
We can thus extract $p_{\rm eff}$ as a function of the filling fraction $f$ and
$\lambda/d$ from our data, and compare with our estimate of $s_{\rm eff}$ and
the relation (\ref{peff}).

Apart from comparing the effective glass exponent measured in our simulations
with experimental data, we may also predict certain features of low-temperature
magnetization relaxation experiments that have been used to locate a possible 
Mott insulating phase. \cite{beauchamp95,nowak96}  
As mentioned in Sec.~\ref{intro_sec}, these experiments identify a dip in the
magnetization relaxation as the signature of a possible Mott insulator. 
Assuming that the relevant relaxation times are determined by variable-range
hopping energy barriers 
$\ln(t/t_0)\approx U(J)/T = (2E_k/T) (J_0/J)^{p_{\rm eff}}$, one finds for the
magnetization relaxation rate $S$ \cite{blatter94,leo95,wengel97}
\begin{equation}
S \equiv - \frac{{\rm d} M}{{\rm d} \ln t} = 
-\frac{{\rm d}\ln J}{{\rm d} \ln t} = \frac{T}{p_{\rm eff}\, U(J)} \ ,
\end{equation}
and hence $S \sim 1/p_{\rm eff}(f)$. 
Previous studies of the magnetization relaxation in the Mott insulator regime
neglected this dependence of $S$ on $p_{\rm eff}$, and rather investigated
quantum fluctuations on the Mott insulator transition, \cite{leo95} which may
well play an important role in addition. 
Here, with our purely classical description, we argue that there may be 
some interesting physics in the behavior of $p_{\rm eff}^{-1}$, which may
already explain recent measurements of the relaxation rate near the expected
Mott insulator transition.
Thus, from plotting $p_{\rm eff}^{-1}$ vs $f$, we can at least qualitatively
compare our results with these experimental data.

\section{Data Analysis} \label{databg_sec}

In this section we present a detailed discussion of our simulation data. 
In the first part, we shall only use uniform, randomly distributed disorder,
and discuss the properties of the Bose glass phase as a function of $\lambda/d$
and $f$.
More specifically, we shall use $\lambda/d=1/4,1/2,1,5$ and $0.3\le f \le 3$. 
The main focus here lies on the question whether there exists a Mott insulating
phase at $f=1$, and if and when it is destroyed, as the interaction strength 
and range between vortices is increased. 
We will also make contact between a number of experimental results and our
simulations.
In the second part, we shall in addition investigate a more correlated spatial
disorder distribution, which only slightly differs from a triangular lattice. 
Here we want to investigate whether the Mott insulator can even exist at large
ratios $\lambda/d$.

Before we start our analysis we would like to discuss finite-size effects. 
First, we want to point out that we do not perform a genuine finite-size 
scaling analysis of our data in this work. 
Rather, we intend to see qualitatively and semi-quantitatively whether the Mott
insulating phase may exist within the Bose glass phase, and if it is easily
destabilized by strong long-range repulsive interactions. 

One may worry, however, about two issues in our simulations:
(i) There may remain artifacts from the underlying grid structure, which we
    employ to make interstitial positions for the vortices available. 
    We have tried to render the grid finer by going from $N=1600$ to $N=3600$ 
    lattice points, keeping $N_D=100$ fixed. 
    However, we did not observe any significant lattice dependence in our
    measurements of $\mu$ or the total energy $E$ in this case. 
    Hence, we believe that our results should be largely insensitive to the
    underlying lattice representation and provide a fair approximation to a
    more realistic continuum description. 
    We therefore used $N=1600$ in most of our simulations presented here.
(ii) One may worry that the number of defects is too small. 
    To check for strong finite-size effects, we therefore chose $N=3600$ and
    increased $N_D$ from 36 to 100 and 225. 
    We did not detect any significant finite-size effects, as was also observed
    in the simulations of a continuous defect distribution. \cite{taeuber95} 
    Also, we were able to make contact with Ref.~[\onlinecite{taeuber95}] by
    choosing $f=0.2$ and $f=0.4$. 
    We could reproduce the previous results in systems of similar sizes, e.g., 
    the values of the chemical potential $\mu$ and the effective exponent 
    $p_{\rm eff}$. 
    Hence, we believe that for the purpose of our largely semi-quantitative 
    analysis we can neglect artifacts stemming from too small lattice sizes or 
    the underlying grid structure.

\subsection{Randomly Distributed Disorder} \label{unidis_sec}

\subsubsection{Strongly vs weakly pinned Bose glass} \label{strwbg_ssec}

We start our data discussion by asking the question of how many flux lines 
become depinned as a result of the vortex interactions, depending on the 
filling fraction $f$ and the interaction range $\lambda/d$. 
As noted in the introduction, within the Bose glass phase one may discriminate 
between regimes of strong and weak pinning (strongly and weakly pinned Bose 
glass, SBG and WBG, respectively), separated by a crossover line. 
The SBG is characterized by the fact that most vortices are pinned to columnar
defects, i.e., {\em single}-vortex pinning is the dominant mechanism. 
In the WBG, on the other hand, only a fraction of lines is pinned by the 
defects, while many other vortices are held in place owing to the repulsive 
interaction exerted from other vortices. 
In this situation, one also speaks of {\em vortex bundle} pinning, i.e., an 
arrangement of flux lines bound to defects effectively form a cage for other
vortices, and pin these through their mutual interaction. 
\cite{nelson92,leo95,larkin95}  
For $\lambda/d \ll 1$ the crossover between the two regimes is expected to
occur at $B\sim B_\Phi$. \cite{nelson92}
Once interactions become strong, however, the WBG should appear well below
$B_\phi$.

\begin{figure}[t]
\centerline{\epsfxsize 7.6cm \epsfbox{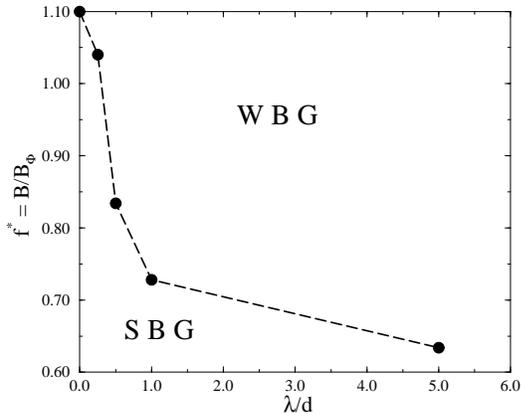}}
\caption{The filling fraction $f^*$, at which 10 \% of the vortices have left
  the defect positions, as a function of $\lambda/d$. The resulting dashed line
  heuristically serves as a crossover line discriminating between the strongly
  pinned Bose glass (SBG), and the weakly pinned Bose glass (WBG), 
  respectively.}
\label{occup_fig}
\end{figure}

Let us define the filling fraction $f^*$, at which 10 \% of the vortices are
depinned in equilibrium as a result of the repulsive interactions. 
With $f^*$ we tentatively identify the crossover line marking the change from 
SBG to WBG in our simulations, as a function of $\lambda/d$.
This line may thus be compared with the crossover line $B^*(T)$ as estimated in
Ref.~[\onlinecite{nelson92}], and experimental results, see 
Refs.~[\onlinecite{krusin96,baumann96}].
Of course, the specific criteria that are used to characterize the boundary
between the SBG and WBG regimes are somewhat arbitrary, but the qualititative
features of $B^*(T)$ should not depend too much on the exact manner this 
crossover line is defined.
Figure~\ref{occup_fig} shows a plot of $f^*$ vs $\lambda/d$. 
One observes a rather strong dependence of this line on $\lambda/d$, which is
shifted well below $B_\Phi$, as soon as the interaction range reaches 
$\lambda \simeq d$. 
Only for $\lambda/d\le 1/4$ does the curve $f^*(\lambda/d)$ remain above $f=1$.
This fact is already a first indication that the Mott insulator does not exist
for $\lambda/d>1/4$, since it is defined by the localization of {\em all} 
vortices on the defects. 
This is, however, impossible, if already 10 \% of the vortices are depinned as
a result of their mutual repulsion at $f^* < 1$. 
Notice that for $\lambda/d=0$, $f^*=1.1$ by definition.

\subsubsection{Site-energy distribution at the matching field}

In order to detect the appearance or disappearance of the Mott insulator more 
clearly, we now turn to the analysis of the (interacting) single-particle 
density of states $g(\epsilon)$ at $f=1$. 
Remember that by density of states we refer to the distribution of pinning 
energies in the flux-line picture, i.e., a histogram of the values of
$\epsilon_i=\sum_{j\neq i} n_i n_j V(r_{ij}) + t_i$ for $k$ realizations of the
disorder distribution.
For long-range interactions one expects a soft ``Coulomb'' gap in $g(\epsilon)$
at the chemical potential $\mu$, whereas for short-range interactions a hard
gap $\Delta$ in the distribution of pinning energies should appear, indicating
the presence of the Mott insulating phase.

\begin{figure}[t]
\centerline{\epsfxsize 8cm \epsfbox{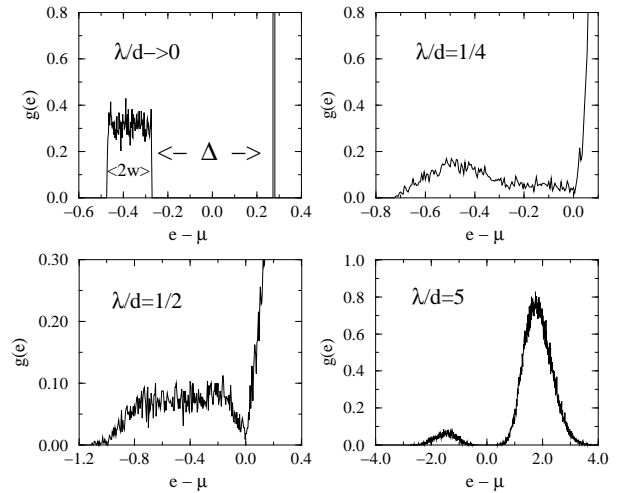}}
\caption{Distribution of (interacting) pinning energies $g(\epsilon)$ vs
  $\epsilon - \mu$ at $f=1$ for four different interaction strengths.  
  From top left to bottom right we have $\lambda/d=0,1/4,1/2,5$. 
  All plots include the data of 50 different realizations of the disorder. 
  In the case of $\lambda/d\to 0$, also the hard gap $\Delta$ separating
  occupied from unoccupied states, and the width of the random site energy
  distribution $2w$ are indicated.} 
\label{dos_fig}
\end{figure}

In Fig.~\ref{dos_fig} we show four different plots of $g(\epsilon)$ vs
$\epsilon - \mu$ for different interaction regimes. 
Let us start at the top left of Fig.~\ref{dos_fig}. 
In the limit $\lambda/d\to 0$ one expects a hard gap in the distribution of
pinning energies, characteristic of the Mott insulator phase. 
This can be seen nicely in this plot, with a gap in $g(\epsilon)$ of width
$\Delta=\langle U_k\rangle -w$ separating the occupied states at 
$e=-\langle U_k \rangle\pm w$ from the empty states at $e=0$, with the chemical
potential $\mu=-(\langle U_k \rangle-w)/2$ (by definition) located in the 
center. 

Let us now study whether the Mott insulator survives the introduction of fairly
weak repulsive interactions.  
The smallest interaction range we can use on a lattice with $N=1600$ and
$N_D=100$, hence $d=\sqrt{N/N_D}=4$, is $\lambda/d=1/4$. \cite{footnote1} 
The top right of Fig.~\ref{dos_fig} depicts $g(\epsilon)$ for $\lambda/d=1/4$.
We observe a fairly flat density of states for the occupied sites, being 
somewhat depleted at $e\lesssim \mu$ and rises sharply for $e > \mu$. 
However, the fact that all states below $\mu$ are {\em continuously} filled, 
implies that even rather short-range interactions suffice to destabilize the 
Mott insulator as a distinct thermodynamic phase, for its unique signature, 
namely the hard gap in the site-energy distribution, has vanished.
Simultaneously, we observe that already 4 \% of the vortices (in an average 
over 50 realizations of the disorder) have left the defect sites, and moved to
energetically more favorable interstitial sites.
We have also looked at $\lambda/d \approx 1/8$ by increasing $N$ to 3600 and
using only $N_D=56$, and hence $d\approx 8$. 
Yet even in this case we found some states in the vicinity of $\mu$ and
approximately 2 \% of vortices on interstitial sites. 
This result strongly indicates that for the true Mott insulator to appear in
the presence of a random, uniform disorder distribution, a necessary condition
is $\lambda\ll d$.
This incompressible phase may, therefore, never be truly realized in e
xperiments, where usually $\lambda\gtrsim d$. 
We speculate, though, that for a more correlated disorder distribution the Mott
insulator may persist to considerably larger interaction ranges; this
suggestion will be further investigated in Sec.~\ref{corrdis_ssec}.

We now turn to the two bottom graphs of Fig.~\ref{dos_fig}.
For $\lambda/d=1/2$ (left) there is clearly no gap, but at  
$\epsilon - \mu\simeq 0$ a narrow, {\em soft} gap begins to appear in the
distribution of pinning energies $g(\epsilon)$. 
In the bottom right figure for fairly long-range interactions $\lambda/d=5$, a
wide soft gap of the form $g(\epsilon) \sim |\epsilon - \mu|^{s_{\rm eff}}$
has fully developed. 
Such a Coulomb gap, though narrower, is also found for $\lambda/d=1$.

\begin{figure}[t]
\centerline{\epsfxsize 4.6cm \epsfbox{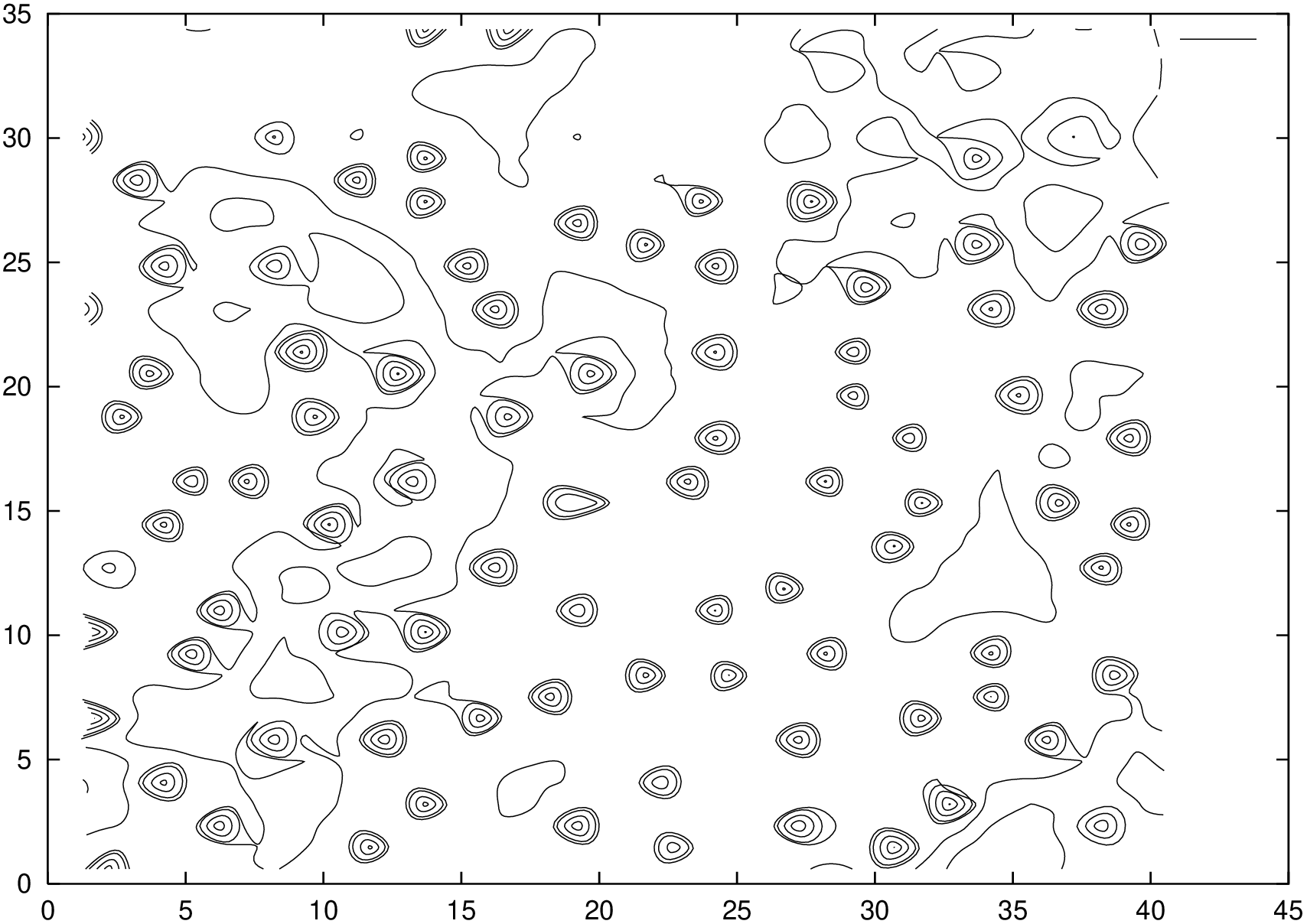}} \par
\centerline{\epsfxsize 4.6cm \epsfbox{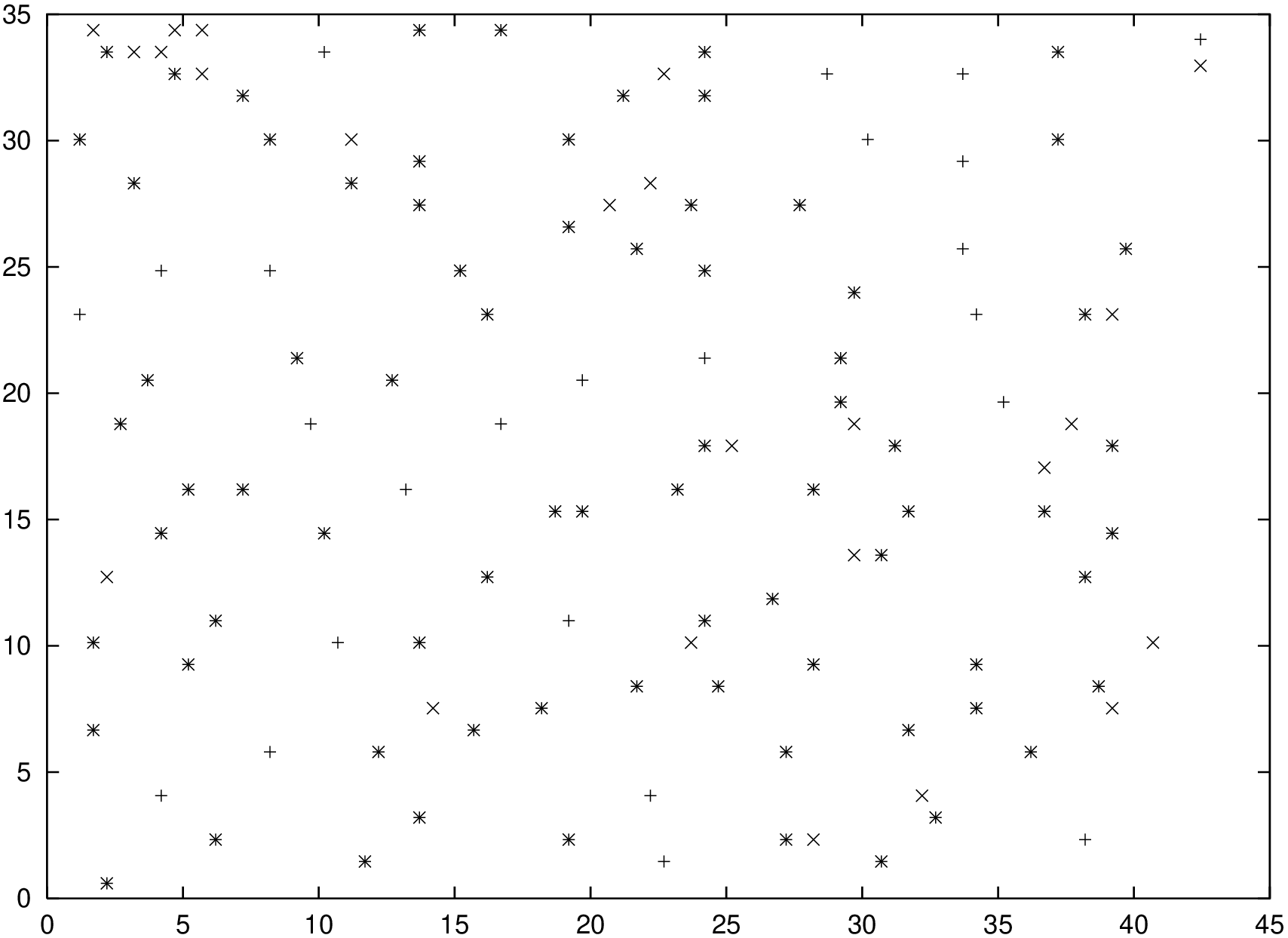}}
\caption{Energy contour plot (top) for $f=1$ and $\lambda/d=5$ for the
  distribution of pinning sites ($\times$) and the equilibrium distribution
  of vortices ($+$). Vortices occupying defects are marked by a star.
  Notice that even vortices that are not bound to defects are located at sites
  characterized by deep energy minima, see, e.g., the particles at coordinates
  $(x=4,y=4)$, $(22,4)$ or $(13,16)$.}
\label{contour_fig}
\end{figure}

We note that in this situation of strong repulsive interactions and large 
filling, the effect of the pinning sites themselves has become rather weak.
The vortices are largely held in place by the cages formed by the other flux 
lines, and the overall structure should be rather unstable against a collective
drift motion, when an external current is applied.
Quite obviously, we are now well in the weakly pinned Bose glass regime.
In Fig.~\ref{contour_fig}, we display a typical energy contour plot of an
equilibrium configuration at $f=1$ and $\lambda/d=5$. 
The top figure depicts the energy contour, where many circular lines indicate
deep energetic minima, corresponding to the defect/vortex configuration of the
bottom plot. \cite{picrem}
For this case of fairly long-range interaction, the energetically favorable 
sites include both the defect and high-symmetry interstitial positions. 
We observe that pinning via the repulsive forces exerted by neighbors, which
are still attached to defects, is very effective. 
As can be seen in Fig.~\ref{contour_fig}, the potential minima are roughly
equally wide and deep, irrespective of the flux line occupying a defect or an
interstitial site. 

In summary, by analyzing the density of states at $f=1$ we find that even for
short-range interactions no distinctive Mott insulator phase appears. 
It becomes rapidly destabilized by the repulsive forces between flux lines, and
only for $\lambda/d\to 0$ do we observe a finite hard gap in the density of 
states near $\mu$, i.e., a Mott insulator.  
In the following, we shall investigate if the measurement of other observables 
corroborates this picture which we obtained by the analysis of the density of 
states.
Furthermore, we discuss the qualititive explanation of recent experiments that
identified (albeit rather broad) features near the matching field.

\subsubsection{Bulk modulus and reversible magnetization}

Let us next turn to the determination of the bulk modulus $c_{11}$ as a 
function of the filling fraction $f$, while keeping $\lambda/d$ fixed.  
A divergence of $c_{11}$ at $f=1$ would indicate the appearance of the Mott
insulator, as an incompressible phase ($\tilde{\kappa}=1/c_{11}=0$).
In order to obtain $c_{11}$ we have measured the chemical potential $\mu$
defined in Eq.~(\ref{mu_eq}) as a function of $f$, i.e., the $H(B)$ curve. 
From this we obtain the bulk modulus by $c_{11}=\partial \mu/\partial f$, the
derivative being of course replaced by a discrete difference when analyzing our
data. 
As we have seen in our study of the density of states, for $\lambda\to 0$ a
hard gap of size $\Delta$ appears. 
In this case $\mu$ jumps from $-\Delta$ to $0$ as $B$ is increased from 
$B_\Phi$ to $B_\Phi$ plus one additional flux quantum. 
Thus, a divergence of $c_{11}$ is trivially connected to a gap in the density
of states. 

\begin{figure}[t]
\centerline{\epsfxsize 8cm \epsfbox{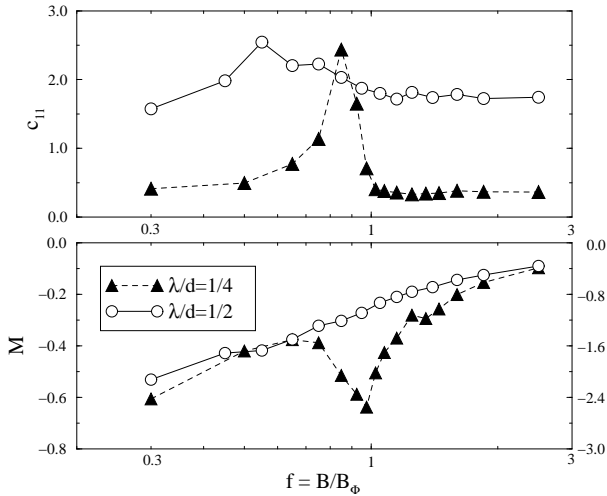}}
\caption{Top: Bulk modulus $c_{11}$ vs $f$ on a log-linear plot. 
  One observes a strong cusp at about $f=0.85$ for $\lambda/d=1/4$ (triangles).
  For $\lambda/d=1/2$ (open circles), the cusp is much less pronounced, and 
  shifted   to a lower filling fraction $f\approx 0.55$. 
  Bottom: Reversible magnetization $M$ vs $f$ on a log-linear plot. 
  For $\lambda/d=1/4$ there is a marked dip in the magnetization at 
  $f\approx 1$, whereas this structure is almost completely absent for
  $\lambda/d=1/2$, with only a shallow minimum at $f\approx 0.6$ remaining.} 
\label{k+m_fig}
\end{figure}

Fig.~\ref{k+m_fig} (top) depicts the bulk modulus vs filling fraction on a 
log-linear plot.
One observes that no divergence appears at $f=1$ for $\lambda/d=1/4$, yet a
quite pronounced cusp occurs in $c_{11}$, shifted downwards to $f\approx 0.85$.
For an even larger interaction range ($\lambda/d=1/2$) the cusp in $\mu$ is
displaced to even smaller $f \approx 0.55$ and less marked; for $\lambda\ge d$,
the bulk modulus becomes essentially constant over the entire range of $f$.
In both cases depicted here, the location of the peak in $c_{11}$ coincides
with that value of $f$ at which a sizeable number of vortices actually leaves
the defect sites.
The fact that the average bulk modulus is enhanced for larger $\lambda$ is
simply an energetic effect, since a vortex entering the system has to overcome
a higher energy barrier due to the interactions.
The behavior of $c_{11}$ confirms again that no true Mott insulator exists even
for comparatively short-range interactions $\lambda/d=1/4$ or $1/2$.
We do observe, though, a distinctive ``lock-in'' structure for $\lambda/d < 1$,
visible as a cusp in $c_{11}$.
This feature, however, completely disappears for $\lambda/d\ge 1$.  

In order to make contact with experiments, we measured the total energy $G$,
which yields the reversible magnetization via the thermodynamic relation 
$M=-\partial G/\partial B \sim -N_V^{-1}\partial G/\partial f$.
Fig.~\ref{k+m_fig} (bottom) depicts $M$ as a function of $f$ on a log-linear
plot for $\lambda/d=1/4$ and $\lambda/d=1/2$, respectively.
The data for $\lambda/d=1/4$ show a pronounced minimum in $M$ at $f\approx 1$,
embedded in a slow logarithmic growth as $f$ increases.
The second plot for $\lambda/d=1/2$ hardly displays any structure aside from a
shallow dip near $f\approx 0.6$.
This feature is completely absent for $\lambda/d\ge 1$, where only the $\ln f$
increase remains, resembling the magnetization curve of an unirradiated
superconductor. \cite{tinkham75,li96,vdbeek96,bulaevskii96} 
The observed behavior of $M$ at $\lambda/d=1/4$ qualitatively agrees very well
with recent measurements performed by van der Beek {\em et al.} in an
irradiated BSCCO crystal, \cite{vdbeek96} where (at $T\approx T_1$) a
pronounced dip in $M$ centered near $B=B_\Phi$ was found. 
The disappearance of this minimum in the experiments as $T$ is increased
towards the transition temperature may be at least partially due to the
divergence of the London penetration depth $\lambda(T)$, which we capture in
our data for large $\lambda$, as well.
In addition, the entropic renormalizations studied in
Ref.~\onlinecite{bulaevskii96} will probably play a significant role.  
Our simulations cannot explain, however, the magnetization minima found at 
$f > 1$ in BSCCO tapes. \cite{li96} 
We will return to this issue in Sec.~\ref{conclusion}.

We would like to point out that the maximum in $c_{11}$ occurs at {\em lower}
value of $f$ than the minimum in $M$ (compare, e.g., our data for 
$\lambda/d=1/4$), which can be understood as follows 
(see Sec.~\ref{rmag_ssec}).
Starting from the relation $M = B - H$ and $\mu \sim H - H_{c_1}$, the bulk
modulus may be written as 
$c_{11} \sim \partial H/\partial B = 1 - \partial M/\partial B$; thus the 
maximum of $c_{11}$ occurs at the location of the steepest negative slope in 
$M(B)$, which has to be at a smaller $B$ than the minimum in $M$. 
Only in the ``true'' Mott insulator phase will the singularities in both
$c_{11}$ and $M$ coincide precisely at $B=B_\Phi$. 

In summary, the measurements of the bulk modulus and the reversible
magnetization confirm the scenario that no Mott insulator phase exists at
finite $\lambda/d$. 
There are, however, interesting ``lock--in'' structures present at
$\lambda/d\le 1$, which qualitatively agree with recent experimental findings. 
We interpret these as {\em remnants} of the Mott insulator phase, but not as
true signatures of this distinctive thermodynamic phase itself.

\subsubsection{I-V characteristics and magnetization relaxation}

In the last section of our data analysis we now turn to the transport
properties of our model, which can be evaluated from the single-particle 
density of states as described in Sec.~\ref{iv+mag_ssec}. 
The calculation of the effective transport exponent $p_{\rm eff}$ by such means
was first used in simulations in Ref.~[\onlinecite{taeuber95}], where small
filling fractions only were analyzed and thus no ``interstitial'' grid was used
as in the present study, but rather a continuous distribution of defect sites. 
Thus, each vortex was always pinned by an energetically favorable defect site
(single-vortex pinning) and hence the equilibrium configurations should be
stable against thermal fluctuations or small currents (strongly pinned Bose
glass regime). 
In our case, where we want to analyze $f\gtrsim 1$ as well, we need to see
whether our equilibrium configurations are sufficiently stable against such
perturbations.
But as the above energy-contour plot (Fig.~\ref{contour_fig}) demonstrates, the
potential minima for the vortices both at defect and at interstitial positions
are about equally deep, and generally well-separated from each other.
This gives us confidence that the single-particle density of states may indeed
be used to infer low-current transport properties in the variable-range hopping
regime. 
It also justifies {\it a posteriori} the quality of our grid description of the
problem: The most important low-energy interstitial sites are correctly taken
into account.

We have determined the effective glass exponent defined by the equation
\begin{equation}
  E = \rho_0 J \exp[-(U_c/T) (J_1/J)^{p_{\rm eff}}] \ ,
\end{equation}
via employing the method described in Sec.~\ref{iv+mag_ssec} in the range 
$0.5\le f \le 3$ for $\lambda/d=1/2,1,5$. 
We were not able to extract $p_{\rm eff}$ from the density of states for
$\lambda/d=1/4$, since the absence of even a small Coulomb gap (cf. 
Fig.~\ref{dos_fig}, top right) renders the definition of $s_{\rm eff}$ 
impossible. 
Our findings are that for interaction ranges $\lambda/d=1,5$ we observe a slow,
monotonous rise in $p_{\rm eff}$ as $f$ is increased: 
For $\lambda/d=1$ from $p_{\rm eff}(f=0.5)\approx 0.54$ to 
$p_{\rm eff}(f=3)\approx 0.74$, and for $\lambda/d=5$ from 
$p_{\rm eff}(f=0.5)\approx 0.7$ to $p_{\rm eff}(f=3)\approx 0.8$. 
It is interesting to compare these results with the earlier simulations at low
filling fractions. \cite{taeuber95} 
For instance, there at $f=0.1$ and $\lambda/d=1$ an effective exponent of 
$p_{\rm eff}\approx 0.5$ was found, which then {\em decreased} almost to the
non-interacting Mott value $p_{\rm eff}=1/3$, as $f$ was increased towards
$0.8$. 
This, however, was due to the fact that the system had to accommodate with the
underlying randomness, since no favorable interstitial sites were available. 
In our case, with these interstitial sites being accessible, the disorder
effects are screened by the interactions as the vortex density is increased,
which leads to {\em stronger} correlations, and presumably more mean-field like
behavior ($p_{\rm eff}\to 1$), because of the more uniform spatial distribution
of flux lines.
Thus, our simulations are presumably more realistic for higher filling 
fractions, and the tendency of increasing transport exponents with larger 
magnetic fields is experimentally confirmed in recent I-V measurements on YBCO.
\cite{baumann96} 
In these experiments, a continuous increase of $p_{\rm eff}$ with $B$ from 
$p_{\rm eff}(f\sim 10^{-2})=1/3$ to $p_{\rm eff}(f\sim 1) \approx 1$ was found.

\begin{figure}[t]
\centerline{\epsfxsize 8cm \epsfbox{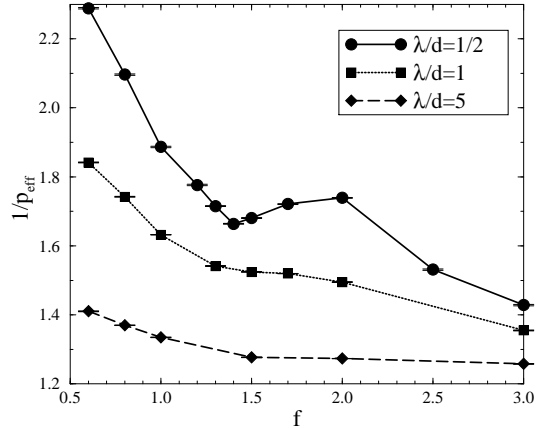}}
\caption{Inverse transport exponent $1/p_{\rm eff}$ vs $f$ for 
  $\lambda/d = 1/2,1,5$. For $\lambda/d = 1/2$, the maximum in $p_{\rm eff}$
  is located at $B \approx 1.4 B_\Phi$.}
\label{peff_fig}
\end{figure}

We plot our data for $p_{\rm eff}^{-1}$ in Fig.~(\ref{peff_fig}). 
For $\lambda/d=1/2$ we find a marked minimum in $p_{\rm eff}^{-1}$ at
$f\approx 1.4$, which is presumably due to a delicate interplay of disorder and
interaction effects. 
Assuming that relaxation times are determined by the variable-range hopping 
energy barriers, we found in Sec.~\ref{iv+mag_ssec} that the magnetization 
relaxation rate $S \sim p_{\rm eff}^{-1}(f)$. 
Thus, the minimum we find in $p_{\rm eff}^{-1}$ for $\lambda/d=1/2$ might
explain low-temperature relaxation experiments on YBCO samples, 
\cite{beauchamp95} as well as on Tl-compounds, \cite{nowak96} for which a 
marked minimum in $S(B)$ at $B\approx 1.4 B_\Phi$ wa reported.  
With the knowledge we have gathered from the above measurements, we would argue
that the minimum detected in the magnetization relaxation experiments is 
{\em not really} a signature of a true Mott insulating phase, but rather a 
remnant of the Mott insulator in the weakly pinned Bose glass near $f=1$.

In summary, our simulation data suggest that in a sample with a uniform random
distribution of columnar defects, a Mott insulating phase is not observable as
a consequence of the repulsive interactions between vortices. 
It may, however, persist up to considerably larger interaction ranges, if the
disorder distribution is chosen in a more correlated manner, e.g., almost 
matching the triangular Abrikosov lattice of a pure type-II superconductor.
\cite{footnote3}
This way the strong repulsion of vortices on defects, which are located very 
closely to each other, is avoided, and the interactions basically lead to an
overall constant upwards shift of all the site energies. 
We are going to investigate this possibility in the following section.

\subsection{Spatially Correlated Disorder Distribution} \label{corrdis_ssec}

In order to investigate the presence of a Mott insulator phase up to larger 
interaction scales in a spatially more correlated disorder potential, we have 
constructed a trial disorder distribution as follows: 
The triangular lattice points with mutual distance $4a$ are looked up on our
underlying grid, and then, with equal probability, the defect is either put on 
the six nearest neighbors of the original point, or it remains on the original 
site. 
Thus, disorder only comes in as a slight distortion of a perfect triangular
defect lattice. 
Related simulations using {\em ideal} triangular and square disorder lattices 
were performed by Anghelache {\em et al.}, \cite{anghelache97} with the
remarkable result that at some values of $f$ and $\lambda / d$, a quadratic
defect array is apparently more effective in pinning the vortices than a
triangular lattice of pins.
We remark, however, that the present Monte Carlo algorithm was specifically
designed for disordered systems, and does not always faithfully reproduce the
correct ground state in ordered structures (typically, regularly ordered
subdomains remain separated by dislocations, which cannot be removed by the
merely local processes allowed for by the algorithm).
Such regular lattices of pins have recently been produced in high-$T_c$
superconductors, and also been investigated experimentally, where the main 
focus lay on the enhancement of the irreversibility line, i.e., $T_c(B)$. 
\cite{baert95}
As for temperatures at least, the pinning efficiency of such regular pin arrays
is much improved, we expect the crossover line between the weakly and strongly 
pinned Bose glass regimes to be drastically shifted upwards.
We also remark that the non-equilibrium properties of such regular arrays under
conditions of strong drive were studied numerically by Reichhardt {\em et al.}
\cite{reichhardt96}

\subsubsection{Strongly vs weakly pinned Bose glass}

We first ask again the question, how many vortices have left their defect
positions as a function of $\lambda/d$? 
It is interesting to note that by looking at the defect occupancy at $f=1$, we
find that even for $\lambda/d=5$ only roughly 1 \% of the vortices have left
the defect positions, preferring an interstitial position. 
At $\lambda/d\le 1$ the vortices are always located on defect sites. 
Thus, the crossover line between the strong and the weak Bose glass which we
have defined pragmatically in Sec.~\ref{strwbg_ssec}, remains above $f=1$ for 
the {\em entire} range of interactions we study here. 
This is in striking contrast to Fig.~\ref{occup_fig}, where we observed a
quickly decreasing crossover line as soon as $\lambda/d$ was increased above
1/4. 
This indicates that a more correlated type of disorder enhances single-vortex
pinning drastically, which has already been verified in experiments.
\cite{baert95}
Moreover, it renders the occurrence of a Mott insulating phase for 
$\lambda/d \lesssim 1$ possible.  

\begin{figure}[t]
\centerline{\epsfxsize 8cm \epsfbox{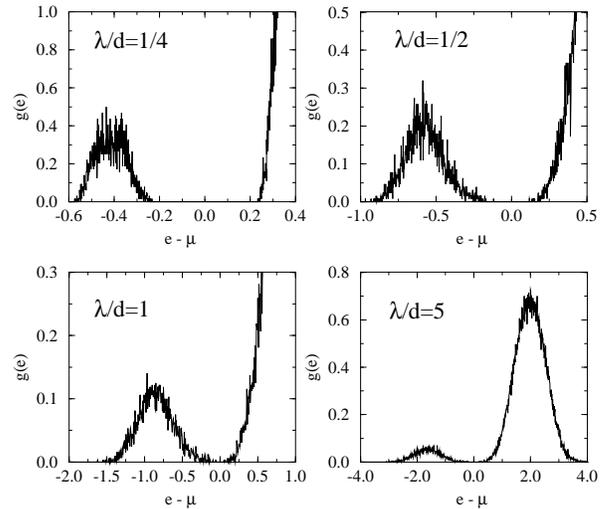}}
\caption{Distribution of pinning energies $g(\epsilon)$ vs $\epsilon - \mu$ for
  four different interaction strengths, using the correlated disorder
  distribution. From top left to bottom right we have $\lambda/d=1/4,1/2,1,5$.}
\label{dos-tri_fig}
\end{figure}

\subsubsection{Site-energy distribution at the matching field}

As in the previous section, we set out to detect a possible Mott insulator by
looking for a hard gap $\Delta$ near the chemical potential in the density of
states $g(\epsilon)$ (distribution of pinning energies). 
Figure~\ref{dos-tri_fig} shows four different plots of the density of states at
$f=1$ for our more correlated disorder distribution. 
Starting with the top left of Fig.~\ref{dos-tri_fig}, we see that at 
$\lambda/d=1/4$ there clearly exists a broad hard gap in $g(\epsilon)$ near the
chemical potential $\mu$. 
All vortices are localized on defects in all 50 realizations of the disorder
and thus a true Mott insulator phase emerges in this interaction regime. 
This is in contrast to the uniform random disorder distribution of the last
section, where for $\lambda/d=1/4$ and even smaller values the Mott insulating
phase was already destroyed (cf. Fig.~\ref{dos_fig}, top right). 

The top right figure with $\lambda/d=1/2$ shows a significantly smaller, but
still finite gap between the occupied low-lying states and the vacant
high-energy states.
Only at $\lambda/d=1$ (left bottom) has the gap closed and a soft Coulomb gap
structure has emerged instead. 
However, all vortices are still bound to defects. 
Only for $\lambda/d=2$ (not depicted here) do the first vortices leave their
defect positions in favor of low-cost interstitial sites. 
Finally, for $\lambda/d=5$ the Coulomb gap structure only insignificantly
differs from the one found for the uniform random disorder distribution (see 
Fig.~\ref{dos_fig} bottom right); the disorder is now effectively screened by 
the long-range interaction between vortices. 

We thus conclude that the Mott insulator phase can survive the introduction of
inter-vortex interactions if the defect structure has a more correlated form,
as studied in our example. 
We find that the Mott insulator can still be found up to $\lambda/d\simeq 1$,
in sharp contrast to the completely random defect distribution, in which case
even for $\lambda/d \simeq 1/8$ no incompressible phase could be identified. 
Let us now see how the Mott insulator phase appears in the measurement of the
bulk modulus and the reversible magnetization.

\subsubsection{Bulk modulus and reversible magnetization}

\begin{figure}[t]
\centerline{\epsfxsize 8cm \epsfbox{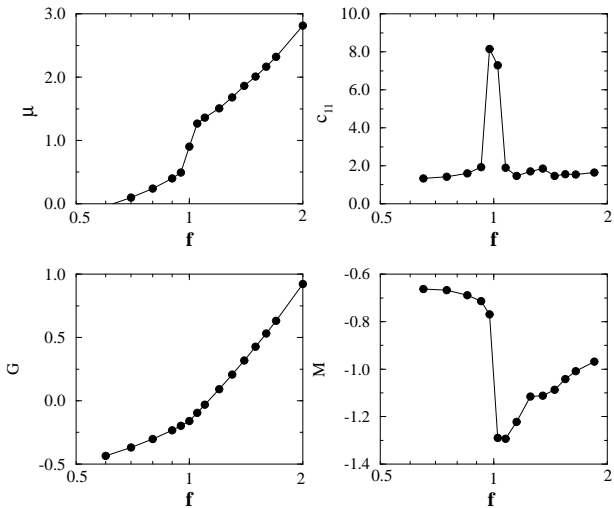}}
\caption{Chemical potential $\mu$ (top left), bulk modulus $c_{11}$ 
  (top right), total energy $G$ (bottom left), and magnetization $M$
  (bottom right) vs filling fraction $f$ on log-linear scales. The
  interaction length scale for all plots is $\lambda/d=1/2$.}
\label{k+m-tri_fig}
\end{figure}

In order to see the effects of the Mott insulator on the bulk modulus and the
reversible magnetization, we consider the interaction regime $\lambda/d=1/2$,
where we still observe a finite gap in the density of states. 
In Fig.~\ref{k+m-tri_fig} we show the chemical potential $\mu$ vs $f$. 
This function displays a distinct jump near $f=1$, slightly smeared out due to
finite-size effects. 
Hence, the bulk modulus $c_{11}=\partial \mu/\partial f$ shows a strong narrow
maximum near $f=1$, which in the limit of an infinitely large system would
presumably yield a $\delta$-function peak. 
Notice that the bulk modulus stays fairly constant below and above $f=1$. 
This behavior in the bulk modulus needs to be contrasted with that of $c_{11}$
for the uniform random disorder distribution and $\lambda/d=1/4$ 
(cf. Fig.~\ref{k+m_fig} top). 
There, at an even shorter interaction scale, only a broad and much flatter 
cusp in $c_{11}$ was found, and its maximum was even shifted to a lower filling
fraction of $f\approx 0.85$. 
This had indicated the absence of a Mott insulator phase, although the system
near $f=0.85$ was clearly fairly stiff with respect to introducing new
vortices. 
Here, however, the narrow and sharp cusp in $c_{11}$ and the gap in the density
of states strongly indicate the presence of the Mott insulator.

The bottom left plot of Fig.~\ref{k+m-tri_fig} shows the total energy $G$ vs
$f$. 
Clearly, there is a marked change in the slope of $G$ as the filling fraction
is tuned through $f=1$. 
Consequently, there is a strong dip in the magnetization 
$M=-\partial G/\partial B$ at $f=1$, as can be seen in the bottom right plot. 
This minimum is much more pronounced and sharper as compared to those found for
$\lambda/d=1/4$ or $1/2$ in the uniform random disorder case (compare the broad
dips in Fig.~\ref{k+m_fig} bottom). 
Also note that both the peak in $c_{11}$ and the dip in $M$ occur at the same
location, namely precisely at $f=1$, as is indicative of the true Mott 
insulator phase (see our discussion in Sec.~\ref{rmag_ssec}). 
To our knowledge, such a sharp dip indicating a real Mott insulating transition
has not been seen in experiments so far. 

Let us just briefly describe the other interaction regimes. 
For $\lambda/d=1/4$ of course the ``divergences'' in $c_{11}$ and $M$ at $f=1$
feature even stronger than in the case depicted in Fig.~\ref{k+m-tri_fig} and,
therefore, also here the Mott insulator is visible. 
$\lambda/d=1$ roughly represents the border line between the occurence of a 
hard gap in the density of states and the soft Coulomb gap. 
The cusp in $c_{11}$ near $f=1$ is still rather narrow, but much less
pronounced, and also the dip in the magnetization is less strong. 
For $\lambda/d=5$, the bulk modulus remains flat over the entire range of the
filling fraction, and the magnetization shows no significant dip.

In summary, we have seen that a more correlated disorder distribution enhances 
single-vortex pinning drastically, and may thus even show a Mott insulating 
transition for moderate interaction ranges. 
The Mott insulator phase should be observable in measurements of the reversible
magnetization, with a sharp dip precisely at $f=1$ being its clear signature.
However, this phase might well be rather unstable with respect to thermal
fluctuations, because correlated vortex hops would be facilitated by the more
regular defect distribution.

\section{Conclusion} \label{conclusion}

The aim of our study was to investigate how the repulsive forces between
vortices affect the properties of the Bose glass phase, and especially the 
proposed existence of a Mott insulator phase, as the magnetic field is varied 
near the matching field $B_\Phi$. 
Specifically, we set out to explain recent experiments which apparently have
found evidence for the Mott insulator transition in irradiated
high-$T_c$-superconductors. 
Our main findings for a uniform random defect distribution, which is
realized in most experiments, are:

\begin{itemize}

\item At the matching field $B_\Phi$, the Mott insulator, being characterized
  by a hard gap in the distribution of pinning energies $g(\epsilon)$ near the
  chemical potential $\mu$, only exists for {\em extremely} short-range 
  interactions $\lambda/d\to 0$. 
  With increasing $\lambda/d$, the gap quickly fills and, therefore, the Mott
  insulator phase as well as the singularities of the accompanying phase 
  transition are destroyed.  
  For long-range interactions $\lambda\gtrsim d$, a soft Coulomb gap described
  by $g(\epsilon)\propto |\epsilon - \mu|^{s_{\rm eff}}$ emerges with an
  effective gap exponent $s_{\rm eff} > 0$. 

\item By measuring the reversible magnetization and the effective I-V exponent
  $p_{\rm eff}$, which we argue is related to the magnetization relaxation
  rate, we are able to explain recent experimental results 
  \cite{vdbeek96,beauchamp95,nowak96} as {\em remnants} of the Mott insulator
  for $\lambda<d$, but {\em not} as true signatures of this thermodynamic phase
  itself. 
  More specifically, the dip in the magnetization at $f\lesssim 1$ and the
  minimum in the relaxation rate $S$ near $B\approx 1.4 B_\phi$ found in the
  experiments can be reproduced in our simulations, albeit with a simultaneous
  absence of a hard gap in the density of states.

\item By tuning $f$ in the interval $[0.5,3]$ for $\lambda/d\ge 1$ we find a
  slowly increasing effective glass exponent $p_{\rm eff}$ with $f$, which is 
  in accord with recent measurement on YBCO films. \cite{baumann96}

\item We can identify a crossover line discriminating between the strongly and 
  weakly pinned Bose glass regimes, respectively. 
  This curve is close to the matching field for $\lambda/d \le 1/4$, but 
  decreases rapidly as the interaction range is increased. 

\end{itemize}

Our simulations cannot explain the experimental finding of Li {\em et al.},
\cite{li96} who observed a minimum in the reversible magnetization at 
$B\approx 1.6 B_\Phi$ in two samples of irradiated BSCCO. 
It is, however, intrinsically difficult to guess the correct value of $B_\Phi$,
which is estimated from knowing the average flux density of the heavy-ion beam 
and some guesses of how effectively these ion columns can create pinning tracks
for the vortices.
Slight deviations of the estimated from the true defect density may therefore
be responsible for this apparent discrepancy of the minimal magnetization and
their $B_\Phi$. \cite{leo97}
In this case there would be no contradiction to our result and that of the
experiments of van der Beek {\em et al}. \cite{vdbeek96}
Of course, we cannot exclude that thermal fluctuations might also play a role,
as these were completely neglected in our algorithm, but these would rather be 
expected to further smoothen any structures in the magnetization.

Can it be understood microscopically why the Mott insulator is destroyed for
even short-range interactions? 
Due to the random distribution of pinning sites, there is a small probability 
in the Poisson distribution that two, three, or even four defect sites are
clustered closely together, say, three defects on a triangle of our grid. 
Although these defects present favorable pinning sites for vortices, their
mutual (though short-range) repulsion may lead to the effect that one vortex
actually leaves in favor of a high-symmetry interstitial site, while the others
remain on the pinning centers. 
Thus, even for short-range interactions, a small but finite percentage of flux
may never sit on defect sites due to the clustering of defects.
Consequently, the Mott insulator, in which all defects are occupied by
vortices, is destroyed.

To cancel this effect we also used a trial disorder distribution, which is
basically a slightly distorted triangular defect lattice. 
We presumed that in this case single-vortex pinning should be enhanced and thus
the Mott insulator should be visible for larger interaction regimes. 
Here our results are as follows: 

\begin{itemize}

\item The crossover line discriminating between the weakly and the strongly 
  pinned Bose glass is drastically shifted upwards. 
  It remains above $f=1$ even for $\lambda/d=5$.

\item At the matching field the Mott insulator can be found up to 
  $\lambda/d \le 1/2$, marked by hard gap in the density of states near the
  chemical potential. 
  The gap fills, and a soft Coulomb gap emerges for $\lambda/d \ge 1$. 

\item For $\lambda \lesssim d$ we find a sharp and narrow peak in the bulk
  modulus at $f=1$; simultaneously, we observe a distinct sharp minimum in the
  magnetization at $B=B_\Phi$. 
  Both signatures are much more pronounced than even for $\lambda/d=1/4$ in the
  uniform random disorder case. 
  The fact that both singularities occur {\em simultaneously} at $f=1$ is 
  another strong indication of the existence of a true Mott insulator. 

\end{itemize}

Our simulation results indicate that in most experimental samples studied so
far, which have hardly any correlations in their disorder distribution, the
Mott insulator is presumably never realized. 
In most high-$T_c$-superconductors the interaction length scale $\lambda$ far
below $T_c$ is of the order 1000 {\AA}. 
The experiments of Li {\em et al.} and van der Beek {\em et al.} used a fairly
high irradiation density, such that the average defect distance was of the
order 500 {\AA}, and thus $\lambda/d \approx 2$. 
Presumably a Mott insulator could be produced with a far weaker radiation dose
such that $d\gg \lambda$. 
Another possibility would be using a more correlated disorder distribution,
like in our simulations, such that repulsive forces from very close vortices
occupying defects can be avoided. 
It would be interesting to investigate both possibilities experimentally. 
Furthermore, the inclusion of thermal fluctuations in the investigation of the
Bose glass properties would be highly desirable.
However, this requires highly efficient algorithms that are able to deal with 
three-dimensional vortex line fluctuations in a disordered environment.

\begin{acknowledgements}

We benefited from discussions with M. Baumann, C.J. van der Beek, J.T. Chalker,
E. Frey, J. K\"otzler, D.R. Nelson, and L. Radzihovsky. 
C.W. wishes to thank A.P. Young for his hospitality and stimulating discussions
at the Physics Department of the University of California, Santa Cruz, where
most of this work has been performed.
U.C.T. acknowledges support from the European Commission through a TMR Marie
Curie Fellowship, contract no. ERB FMBI-CT96-1189, and from the Deutsche
Forschungsgemeinschaft through a Habilitation Fellowship, DFG-Gz. Ta 177/2-1,2.

\end{acknowledgements}


\begin{thebibliography}{99}

\bibitem{blatter94}
  For a recent review, see, e.g., 
  G.~Blatter, M.V.~Feigel'man, V.B.~Geshkenbein, A.I.~Larkin, and V.M.~Vinokur,
  Rev. Mod. Phys. {\bf 66}, 1125 (1994).

\bibitem{nelson88}
  D.R.~Nelson, Phys. Rev. Lett. {\bf 60}, 1973 (1988);
  D.R.~Nelson and H.S.~Seung, Phys. Rev. B {\bf 39}, 9153 (1989). 

\bibitem{fisher91}
  M.P.A.~Fisher, Phys. Rev. Lett. {\bf 62}, 1415 (1989);
  D.S.~Fisher, M.P.A.~Fisher, and D.A.~Huse, Phys. Rev. B {\bf 43}, 130 (1991).

\bibitem{nelson89}
  D.R.~Nelson, J. Stat. Phys. {\bf 57}, 511 (1989). 

\bibitem{lyuksyutov92}
  I.F.~Lyuksyutov, Europhys. Lett. {\bf 20}, 273 (1992).

\bibitem{nelson92}
  D.~R.~Nelson and V.~M.~Vinokur, Phys. Rev. Lett. {\bf 68}, 2398 (1992); 
  Phys. Rev. B {\bf 48}, 13 060 (1993).

\bibitem{fisher89}
  See, e.g., M.P.A.~Fisher, P.B.~Weichman, G.~Grinstein, and D.S.~Fisher, 
  Phys. Rev. B {\bf 40}, 546 (1989), and references therein.

\bibitem{taeuber97}
  For finite $L < \infty$, open boundary conditions are more realistic for a
  flux-line system than the periodic boundary conditions along $z$ required for
  the boson mapping. Some consequences, e.g., the ensuing different finite-size
  corrections in the flux liquid phase, are discussed in:
  U.C.~T\"auber and D.R.~Nelson, Phys. Rep. {\bf 289}, 157 (1997).

\bibitem{shklovskii84}
  See B.I.~Shklovskii and A.L.~Efros, {\em Electronic properties of doped
  semiconductors} (Springer, Berlin 1984); and references therein.

\bibitem{davies82}
  J.H.~Davies, P.A.~Lee, and T.M.~Rice, Phys. Rev. Lett. {\bf 49}, 758 (1982);
  Phys. Rev. B {\bf 29}, 4260 (1984).

\bibitem{moebius92}
  A.~M\"obius, M.~Richter, and B.~Drittler, Phys. Rev. B {\bf 45}, 11568
  (1992).

\bibitem{taeuber95}
  U.C.~T\"auber, H.~Dai, D.R.~Nelson, and C.M.~Lieber, 
  Phys. Rev. Lett. {\bf 74}, 5132 (1995); 
  U.C.~T\"auber and D.~R.~Nelson, Phys. Rev. B {\bf 52}, 16 106 (1995). 

\bibitem{konczykowski95}
  M.~Konczykowski, N.~Chikumoto, V.M.~Vinokur, and M.V.~Feigel'man,
  Phys. Rev. B {\bf 51}, 3957 (1995).

\bibitem{vdbeek95}
  C.J. van der Beek, M.~Konczykowski, and V.M.~Vinokur, 
  Phys. Rev. Lett. {\bf 74}, 1214 (1995).

\bibitem{baumann96}
  M.~Baumann and J.~K\"otzler, private communication (1996).

\bibitem{thompson97}
  J.R.~Thompson, L.~Krusin-Elbaum, L.~Civale, G.~Blatter, and C.~Feild,
  Phys. Rev. Lett. {\bf 78}, 3181 (1997).

\bibitem{wahl95}
  A.~Wahl, V.~Hardy, J.~Provost, C.~Simon, and A.~Buzdin,
  Physica C {\bf 250}, 163 (1995).

\bibitem{leo95}
  L.~Radzihovsky, Phys. Rev. Lett. {\bf 74}, 4919 and 4923 (1995).

\bibitem{bulaevskii96}
  L.~N.~Bulaevskii, V.M.~Vinokur, and M.P.~Maley, 
  Phys. Rev. Lett. {\bf 77}, 936 (1996).  

\bibitem{reichhardt96}
  C.~Reichhardt, C.J.~Olson, J.~Groth, S.~Field, and F.~Nori,
  Phys. Rev. B {\bf 53}, R8898 (1996).

\bibitem{li96}
  Q.~Li, Y.~Fukumoto, Y.~Zhu, M.~Suenaga, T.~Kaneko, K.~Sato, and C.~Simon,
  Phys. Rev. B {\bf 54}, R788 (1996).

\bibitem{vdbeek96}
  C.J. van der Beek, M.~Konczykowski, T.W.~Li, P.H.~Kes, and W.~Benoit, 
  Phys. Rev. B {\bf 54}, R792 (1996).

\bibitem{beauchamp95}
  K.M.~Beauchamp, T.F.~Rosenbaum, U.~Welp, G.W.~Crabtree, and V.M.~Vinokur, 
  Phys. Rev. Lett. {\bf 75}, 3942 (1995). 

\bibitem{nowak96}
  E.R.~Nowak, S.~Anders, H.M.~Jaeger, J.A.~Fendrich, W.K.~Kwok, R.~Mogilevsky,
  and D.G.~Hinks, Phys. Rev. B {\bf 54}, R12 725 (1996). 

\bibitem{wengel97}
  A brief account of these results was already presented in:
  C.~Wengel and U.C.~T\"auber, Phys. Rev. Lett. {\bf 78}, 4845 (1997).

\bibitem{baert95}
  See, e.g., M.~Baert, V.V.~Metlushko, R.~Jonckheere, V.V.~Moshchalkov, and
  Y.~Bruynseraede, Phys. Rev. Lett. {\bf 74}, 3269 (1995);
  J.-Y.~Lin, M.~Gurvitch, S.K.~Tolpygo, A.~Bourdillon, S.Y.~Hou, and
  J.M.~Phillips, Phys. Rev. B {\bf 54}, R12 717 (1996);
  J.I.~Mart\'{\i}n, M.~V\'elez, J.~Nogu\'es. and I.K.~Schuller,
  Phys. Rev. Lett. {\bf 79}, 1929 (1997).

\bibitem{larkin95}
  A.I.~Larkin and V.M.~Vinokur, Phys. Rev. Lett. {\bf 75}, 4666 (1995).

\bibitem{krusin96}
  L.~Krusin-Elbaum, L.~Civale, J.R.~Thompson, and C.~Feild,
  Phys. Rev. B {\bf 53}, 11 744 (1996).

\bibitem{footnote1}
  Using a smaller interaction length $\lambda$ renders the interaction range
  shorter than the underlying grid; this, however, effectively yields
  $\lambda \to 0$ and is thus unphysical.  

\bibitem{picrem}
  The top plot should not be considered above a y-axis value of 30, since due
  to an error of the plot program, the true energy distribution is not 
  represented there.

\bibitem{tinkham75}
  M.~Tinkham, {\em Introduction to Superconductivity} (McGraw-Hill, New York,
  1975).

\bibitem{footnote3}
  Taking a perfect triangular defect lattice would not work at non-zero
  temperatures, since in the quantum-mechanical analogy the localized wave
  functions of the bosons (at $T > 0$) would correspond to delocalized Bloch 
  waves, much like electrons in a perfectly regular potential. 
  In the vortex picture, this would lead to correlated hops of flux lines
  [D.R.~Nelson, private communication (1996)].
 
\bibitem{anghelache97}
  R.~Anghelache, private communication (1997).

\bibitem{leo97}
  L.~Radzihovsky, private communication (1997).

\end{thebibliography}
\end{document}